\documentclass[12pt]{report}
\usepackage{latexsym}
\usepackage{graphicx}
\begin{document}
\newcount\nummer \nummer=0
\def\f#1{\global\advance\nummer by 1 \eqno{(\number\nummer)}
      \global\edef#1{(\number\nummer)}}
\def\Di{\displaystyle}
\def\nn{\nonumber \\}
\def\be{\begin{equation}}
\def\ee{\end{equation}}
\def\ba{\begin{eqnarray}}
\def\ea{\end{eqnarray}}
\def\la{\label}\def\pl{\label}
\def\re{(\ref }
\def\rz#1 {(\ref{#1}) }
\def\el#1 {\label{#1}\end{equation}}
\def\rp#1 {(\ref{#1}) }
\def\i{{\rm i}}
\let\a=\alpha \let\b=\beta \let\g=\gamma \let\d=\delta
\let\e=\varepsilon \let\ep=\epsilon \let\z=\zeta \let\h=\eta \let\th=\theta
\let\dh=\vartheta \let\k=\kappa \let\l=\lambda \let\m=\mu
\let\n=\nu \let\x=\xi \let\p=\pi \let\r=\rho \let\s=\sigma
\let\t=\tau \let\o=\omega \let\c=\chi \let\ps=\psi
\let\ph=\varphi \let\Ph=\phi \let\PH=\Phi \let\Ps=\Psi
\let\O=\Omega \let\S=\Sigma \let\P=\Pi \let\Th=\Theta
\let\L=\Lambda \let\G=\Gamma \let\D=\Delta

\def\w{\wedge}
\def\0{\over } \def\1{\vec } \def\2{{1\over2}} \def\4{{1\over4}}
\def\5{\bar } \def\6{\partial }
\def\7#1{{#1}\llap{/}}
\def\8#1{{\textstyle{#1}}} \def\9#1{{\bf {#1}}}

\def\({\left(} \def\){\right)} \def\<{\langle } \def\>{\rangle }
\def\lb{\left\{} \def\rb{\right\}}
\let\lra=\leftrightarrow \let\LRA=\Leftrightarrow
\let\Ra=\Rightarrow \let\ra=\rightarrow
\def\ul{\underline}

\let\ap=\approx \let\eq=\equiv 
\let\ti=\tilde \let\bl=\biggl \let\br=\biggr
\let\bi=\choose \let\at=\atop \let\mat=\pmatrix
\def\CL{{\cal L}}\def\CX{{\cal X}}\def\CA{{\cal A}}
\def\CF{{\cal F}} \def\CD{{\cal D}} \def\rd{{\rm d}} 
\def\rD{{\rm D}} \def\CH{{\cal H}} \def\CT{{\cal T}} \def\CM{{\cal M}}
\def\CI{{\cal I}} \newcommand{\dR}{\mbox{{\sl I \hspace{-0.8em} R}}} 
\def\CP{{\cal P}}\def\CS{{\cal S}}\def\C{{\cal C}}

\date{May 1994}
\title{Poisson Structure Induced Field Theories and Models of $1+1$
Dimensional Gravity}
 \author{T.\ Strobl} 
\maketitle
\tableofcontents

\chapter{Introduction}

One of the most prominent open problems in theoretical physics is to find some
common understanding of the standard model on the one hand and Einsteins'
theory of gravity on the other hand. Both theories have their own esthetic
appeal, the former because it unified fundamental forces and the latter because
of its geometric interpretation; both theories, furthermore, found sufficient
experimental support within the realm of their validity. However, they are in
conflict with each other: Within general relativity the matter system remains
unquantized, whereas the standard model inherently is a quantum theory. The
attempt to quantize gravity by means of the (perturbative) methods used
successfully in the standard model failed thus far.  This led to various 
alternative approaches.  Beside string theory and noncommutative geometry the
most prominent among these is the search for a consistent {\em
nonperturbative} quantum theory of the coupled Einstein-Yang-Mills-matter
system. Old hopes for success into this direction found some revival due to the
pioneering works of Ashtekar \cite{Ash}, Rovelli and Smolin \cite{Rov}.

To get a better grasp on technical as well as conceptual problems encountered
in this approach the study of the quantization of truncated versions of the
full theory (Bianchi models) or, related to it, of lower dimensional models is
suggestive. Whereas in three space-time dimensions the Einstein-Hilbert action
\be \int d^3x \sqrt{-g}R \el 3D
for gravity is  meaningful (here $R$ denotes the torsionless Ricci scalar
and $g$ the determinant of the metric),  in two space-time dimensions it yields
no field equations, because it is a boundary term: $\int_M d\o(e^a)=\int_{\6 M}
\o(e^a)$, where $\o(e^a)$ is the torsionless spin connection. This led to the
proposal of various other gravity actions in two dimensions.

One of these, proposed by Jackiw and Teitelboim \cite{JT}, has the form \be
L^{JT} \propto \int_M d^2x \sqrt{-g} \Phi (R-\mbox{const}) \,
,\el JT where $\Phi$ is some Lagrange multiplier field. Another
action  studied in two dimensions is
\be L^{R^2}= \int_M d^2x \sqrt{-g} (R^2/16 + \L) \, . \el R2
In contrast to \rp JT it  is purely geometrical. It, however, leads to higher
derivative equations of motion for the metric. Using for \re{R2}) Cartan
variables,  the torsion zero condition does not evolve as an equation of
motion,\footnote{Contrary to what happens in the Palatini formulation of the
four-dimensional Einstein-Hilbert action.} but it has to be implemented  via a
Lagrange multilier or by expressing $\o$ in terms of the zweibein $e^a$.  The
most natural Lagrangian for two-dimensional gravity when using Cartan variables
was proposed by Katanaev and Volovich \cite{KV}:
\be  L^{KV} = \int [-{1 \0 4} d\o \w \ast d\o - {1\0 2\a} De^a \w \ast De_a
+ \L \e] \, . \label{KV} \ee In two space-time dimensions this is the most
general Lagrangian yielding second order differential equations for zweibein
and spin-connection; it is purely geometrical and (but) torsion $De^a$ became
'dynamical'. Another model of pure 2D gravity, gaining much interest recently,
is defined by the  string-inspired action \cite{BH}
\be L^{str} \propto \int_M d^2x \sqrt{-g} \exp (-2 \Phi) [R+
4 g^{\m\n} \6_\m \Phi\6_\n \Phi  -\L]\el string
where $\Phi$ is  the Dilaton field. 

The action (\ref{KV}) has some formal similarity with the one of a Yang-Mills
theory for the Poincar\'e group. For $\L=0$ the only (but decisive) difference is
that the Hodge dual operation is not taken with respect to some background
metric, but with part of the 'Poincar\'e connection', namely the zweibein,
itself. Implementing the torsion zero condition in \re{JT}), on the other hand,
through a Lagrange multiplier $X_a$, the Jackiw-Teitelboim (JT) model  can be
formulated equivalently as a connection flat gauge theory
\cite{Isl} \be L=\int X^iF_i \, ,\el gaugeth where $X^3=\Phi$ and 
the $F_i$ are the components 
of the $so(2,1)$-curvature two-form corresponding to the connection ($a \in
\{1,2\}$)
\be A_a\equiv e_a \quad A_3 \equiv \o \,. \el identi
 Similarly the action \rp 3D was found to be {\em equivalent} to
a $ISO(2,1)$-Chern-Simons gauge theroy \cite{Wit}. Even the Ashtekar
formulation of 4D gravity has some striking similarities (but also
differences!) with a 4D (nonabelian) gauge theory, which are, e.g., the reason
for the successful use of Wilson loops within the gravity theory \cite{Rov}.

Observations such as these and the partial success in finding the quantum
theory for \rp 3D and \rp JT due to the gauge theory formulations led some
people to  reinterpret the vielbein of any gravity theory as the 'missing' part
of a Poincar\'e connection beside the spin (or Lorentz) connection; any gravity
theory becomes a Poincar\'e gauge theory then by an appropriate introduction of
additional auxiliary fields \cite{Gri}.  The flaw in this approach is that in
general the diffeomorphism invariance, which is the main cause for the problems
in the canonical quantization of gravity,  remains still independent of the
Poincar\'e gauge  transformations (in contrast to what happens, e.g.,  with
\re{3D})). This becomes most obvious in the Hamiltonian formulation \cite{Com}.
Still, the question remains: Can one  draw (further) profit from the common
structures of gravity and nonabelian gauge theories?

In two dimensions this question can be answered to the positive. The common
structure between 2D Yang-Mills theories and (at least most)  models of pure 2D
gravity has a name: It is a Poisson structure $P$ in the {\em target space}  of
the theory \cite{9402}.  The first order action for these theories has the
common form\footnote{Independently of us the study of an action equivalent to 
\rp actionform has been proposed also in \cite{Ikeda}.}
\be \int_M A_i \wedge dX^i + \2 P^{ij}(X) A_i \wedge A_j \, . \el actionform  
Here $X^i(x)$ is the map from the space-time or worldsheet manifold $M$ to the
target space $N$,  $P$ is the Poisson tensor defined on the latter 
space, and $A_i$ is a one-form on $M$.  E.g., a Poisson structure linear in $X$
yields  gauge theories of the form  \re{gaugeth}); or the action \re{KV}) can
be reproduced by the choice of a quadratic Poisson structure when integrating
out $X$ and making use of the identification \re{identi}).

The action \rp actionform (and an appropriate extension of it) allows not 
only to study a large 
class of  two-dimensional gravity as well as Yang-Mills 
theories at one  and the same time and footing, the knowledge of having to 
deal with a Poisson structure on the space $N$ suggests also the use of 
otherwise unusual kind of methods. In particular, diffeomorphisms in the 
target space can be used to bring $P$ into some standard form generalizing 
the Darboux form of a nondegenerate $P$. In this way previous lengthy
calculations can be reduced to some lines and the possibility to solve the
incorporated theories on the quantum and classical level for all kind of
different topologies seems close at hand. 

To get a first feeling for the theory as defined in \re{actionform}), 
let us use the field equations of the $A_i$  
to 'integrate them out'  within this action. For simplicity we 
assume that $N \sim \dR^n$ and that $P$ is nondegenerate. $P$ then has an 
inverse $\O$, which is a symplectic two-form on $N$. Up to a multiplicative
factor the action then takes the form \be \int_M \O_{ij} dX^idX^j \, .\el path
This illustrates an important characteristic of the model: Since $\O$ is
closed, \rp path is a Wess-Zumino type action and thus, for finite $n$, 
\rp actionform defines a model  with only a {\em finite} number of (physical)
degrees of freedom.\footnote{Recently I found the reference \cite{Singer},
where the  action \re{path}), arising from \rp
actionform for nondegenerate $P$, was studied from some other perspective 
and shown to be 
equivalent to Witten's topological sigma model on the quantum level 
(cf.\ also \cite{Blau}). 
It will be interesting to further investigate \re{actionform}) 
in view of this connection, also 
for degenerate Poisson structures.}

The organization of this report is as follows: In the next chapter we study (an
extension of) the theory \rp actionform in its own right. To not mix up the
structures defined on $M$ and $N$, we start with a study of Poisson structures
defined on some finite dimensional manifold $N$. In the following section we
then define the action providing details about its symmetry content,
Hamiltonian and BRS formulation, etc. Thereafter we study the classical theory.
Locally its integrability is basically trivial in this formulation. But part of
the field equations are solved also for completely arbitrary topologies of $M$
and $N$. The remaining equations of motion are, furthermore, particularly
simple in an appropriate local coordinate system on $N$. In the concluding
section of this chapter we then come to the quantum theory as defined on
$M=S^1 \times \dR$, such that one may use standard Hamiltonian methods. We
construct {\em all} quantum states. Up to some technicalities of topological
origin to be explained there, the 'physical' wave functionals are basically
functions of a finite number of variables only, as  expected already from
\re{path}).

The third and last chapter focuses on the gravity version of \re{actionform}).
First we find the most general class of models contained in this action which
allows for a gravitational interpretation via \re{identi}).  Restricting
ourselves to a subclass of these, including all torsion-free ones, we show that
locally the metric always can be brought into the 'generalized Schwarz-schild
form' \be g=h(r)(dt)^2 - {(dr)^2 \0 h(r)} \, , \el Schwarz  where {\em any}
function $h$ can be provided by an appropriate choice of the Lagrangian.  We
then solve the equations for the extremals in all 
generality and construct the
universal covering solutions by means of Penrose diagrams. The considerations
are illustrated at  the examples of \re{JT}), \re{R2}), and \re{KV}) (cf.\
Figs.\ 5,6,8). On the quantum level the wave functions depend on one continuous
parameter. Additional discrete labels of the wave functions arise in the case
of a nontrivial causal structure of the classical theory, i.e.\ if $h$ vanishes
at some values of $r$.
 
In the remaining sections we take up three issues which might have their
parallels also in the Ashtekar approach to quantum gravity. Firstly, the
restriction to topologies of the form $\S \times \dR$, characteristic for any
Hamiltonian treatment, is called into question.  By means of the previously
obtained Penrose diagrams we construct {\em all} global solutions for the
example of the Katanaev-Volovich (KV) model \re{KV}) (with Minkowski
signature). The space of these solutions is then compared to the reduced phase
space RPS ($\sim$ space of solutions on $M=S^1 \times \dR$ modulo symmetry
transformations) underlying the quantum theory. On the one hand the numbers of
continuos and discrete parameters  fit nicley, if we strictly stay with the
cylindrical solutions. On the other hand, parts of the RPS are found to
correspond to classical solutions for which some other topology, as, e.g., a
torus with hole, would be  more natural.

Secondly,  \rp actionform stays well-defined also for a configuration
corresponding to a  degenerate metric; furthermore, the Hamiltonian symmetries
identify nondegenerate metrics with degenerate ones. This immitates somewhat
the Ashtekar formulation of four dimensional gravity, which is also nonsingular
at degenerate metrics \cite{Ashbuch}. For the two-dimensional models at hand,
a detailed comparison of the standard Hamiltonian RPS with the one resulting
from dividing out conventional gravity symmetries is possible. It reveals some
inequivalence even after having excluded the nondegenerate solutions. The
solutions identified in the Hamiltonian formulation differ by different kink
number $k$. However, all solutions with $k \neq 0$ turn out to be geodesically
incomplete.

Thirdly, we study the example of $R^2$-gravity coupled to an $SU(2)$-Yang Mills
theory from the conceptual point of view. As in any quantum  theory of gravity
the Dirac observables are  space-time independent and the
Hamiltonian vanishes on physical quantum states.  Strategies to
resolve this apparent 'problem of (space-)time' \cite{Ish}  are
developed at the example of the reparametrization invariant nonrelativistic
particle.  Realizing these strategies in the gravity-Yang-Mills system, one
finds some partial confirmation of them through the fact that a gravity flat
limit reproduces the usual $SU(2)$ quantum dynamics.

\vskip5mm

{\bf Acknowledgement:} 
 \vskip2mm

The progress summarized in this work would have been unimaginable without the
input of two valuable collaborators: Peter Schaller and Thomas Kl\"osch.  Let
me thank them at this point.  Sections 3.2, 3.3, and 3.5 will be published
together with T.\ Kl\"osch \cite{Klo}. Practically all of the rest either will
be or already has been published  together with P.\ Schaller
\cite{9403,9402,p6}. Works paving the way for the present account are further
\cite{All,p2,p5}, beside the works of colleagues of mine at 
my institute and elsewhere \cite{DomLC,Kummeretal,Kat}.  Let me express also
my gratitude to my supervisor Prof.\ W.\ Kummer for encouragement during the
time of the thesis and to Prof.\ W.\ Thirring for his steady interest.

\chapter{Poisson Structure Induced Two Dimensional Field Theories}

\section{Poisson Structures and Symplectic Leaves} 

\label{sec1}

Let $N$ denote a finite dimensional manifold and $\CF(N)$ the space of smooth
functions on it. A Poisson bracket $\{ \cdot, \cdot \}$ on $\CF(N)$ is a
bilinear map $\CF(N) \times \CF(N) \to \CF(N)$ which is skew-symmetric
$\{F,G\}=-\{G,F\}$, obeys the Jacobi identity
\be \{F,\{G,H\}\}+ \{H,\{F,G\}\}+ \{G,\{H,F\}\}=0 \, , \el Jacobi
and fulfills the Leibnitz rule: $\{F,GH\} =\{F,G\}H+G\{F,H\}$.  Due to the
latter requirement and the bilinearity any Poisson bracket can be represented
by a (skew-symmetric) bivector field $P \in \L^2(TN)$:
\be \{F,G\} = P(F,G) = P^{ij}(X) { \6 F(X) \0 \6 X^i } 
{ \6 G(X) \0 \6 X^j }
\, , \el Poi
where we have chosen local coordinates $X^i, \, i = 1, ... ,n$ on $N$.  The
Jacobi identity becomes
\be P^{k[l} P^{ij]},_k =0  \, ,  \el PJacobi
where $[...]$ denotes antisymmetrization and the comma a derivative.  In more
abstract terms, it becomes the vanishing of the Schouten-Nijenhuis bracket  of
$P$ with itself. The latter bracket is a natural (graded) extension of the
Poisson bracket (resp.\ Lie bracket) to $\L(TN)=
\sum_{l=0}^{n} \L^l(TN)$ (cf., e.g., \cite{3Mad}, \cite{Hen}).  

A Poisson structure $P$ is more general than a symplectic one since $P$ need
not be nondegenerate. Locally any Poisson structure $P \in \L^2(TN)$ is
characterized only by $n$, the dimension of the underlying manifold $N$, as
well as the  (local) dimension $k$ of the kernel of $P$. An exception to this
occurs for 'singular points' in  $N$ which are not part of any  neighborhood
with constant $dim \, ker \,  P$.

Let us expand on this: The insertion of any one-form $e \in T^\ast M$ not in
the kernel of $P$ provides  a vector field; the latter is called (locally)
Hamiltonian, if (locally) $e=dF$ for some function $F \in \CF(N)$.  As a
consequence of the Jacobi identity \re{PJacobi}), the set of locally Hamiltonian
vector fields is in involution.  Thus, according to the  Frobenius theorem,
locally they generate an integral surface $S$ through any point $p \in N$ and
it is always possible to introduce local coordinates $X^i=(X^A,X^\a)$, $A=1,
..., k$, $\a = 1, ..., s=n-k$  in $N$ such that $S$ can be described by
$X^A=\mbox{const}$.  $dX^A$ span the $k$-dimensional kernel of $P(p)$ then and
the restriction of $P$ onto $S$, $P\vert_S$, is nondegenrate.  Since  the
restriction of a Poisson bracket to functions on a submanifold yields again a
Poisson bracket, the inverse of $P\vert_S$  is a symplectic (i.e.\ closed and
nondegenrate) two-form $\O \in \L^2 T^\ast S$.  By means of an appropriate
change of variables $X^\a$, it is now always possible to locally bring $\O$
into Darboux form (cf., e.g., \cite{Thi1,Wood}) simultanously on any
of the symplectic leaves $S$.

In the generic case $dX^A$ spans the kernel of $P$ in a neighborhood of $p$ so
that locally the Poisson tensor takes the form: $P=\sum_{l=1}^{s/2} {\6 \0 \6
q^l} \wedge {\6 \0 \6 p_l}$.  (Of course $s$ is an even integer as $\det \O
\equiv \det \O^T = (-1)^s \det \O \neq 0$.) 
In the following we shall call any coordinate system in which $P$ takes this simple form a
Casimir-Darboux coordinate system $(X^A,X^I) \equiv (X^A,q^\cdot,p_\cdot)$. 
 The case of a singular point is included
\cite{Wein}, \cite{3Mad}, if one adds to the previous expression for $P$ the
term $U=(1/2)\sum_{u,v=1}^{n-s} U^{uv} {\6 \0 \6 X^u} \wedge {\6 \0 \6 X^v}$,
where  $U$ depends only on the coordinates $X^u$; $U$  is a Poisson structure
by itself and vanishes at the considered point $p$.

Vice versa, it is obvious that {\em any} choice of a (generalized) foliation of
a manifold $N$ into symplectic leaves, such that the symplectic two-form $\O$
on each of them can be extended into a smooth two-form $\ti \O$ on $N$,
defines a Poisson structure $P$ on $N$.

If there is an additional structure defined on $N$, giving rise to a referred
coordinate system, Poisson strucures identified in the above considerations may
need to be distinguished. For instance the manifold $N$ could be a linear space
such that only linear transformations on $X$ are admissible.  We then find
that, the choice of a Poisson structure $P$ linear in these coordinates,
$P^{ij}=f^{ij}{}_kX^k$, is equivalent to the specification of a Lie algebra
(since \rp PJacobi reduces to the Jacobi identity for the coefficients
$f^{ij}{}_k$), whereas a polynomial $P$ yields a $W$-algebra. Another instance
where some coordinates are distinguished on $N$ is the case where $N$ is some
Lie group $G$.  A Poisson structure satisfying some specific compatibility
condition with respect to the group multiplication on $G$ is called a Lie
Poisson structure, the current interest in which stems from the fact that it
provides the classical limit of a quantum group \cite{Dri2}. In the context of
the gravity models considered in chapter \ref{cha3} $N$ will play the role of
a target space (cf.\ also the Introduction).  In this case the additional input to
the otherwise $N$-diffeomorphism invariant theory will stem from the {\em
interpretation} of specific coordiantes as gravity variables; e.g.\ within the
Katanaev-Volovich model
\re{KV}) the coordinate $X^3$ will play the role of 
the curvature scalar on the underlying world-sheet or space-time manifold $M$,
as an indirect consequence of the identification \re{identi}).

To obtain the most general solution to
\rp PJacobi in terms of  explicit functions on $N$, we only need to apply a
general diffeomorphism $X \to Y$ to the 'Casimir-Darboux form' of the Poisson
structure  obtained above:\footnote{Let us here only be interested in the
generic local shape of $P$, i.e.\ in $P$ in the vicinity of nonsingular
points.}
\be
\begin{array}{c} P^{ij}(Y)= \left( {\Di \6 X(Y) \0 \Di \6 Y} \right)^{-1} 
\left( \begin{array}{cc} 0 & 0 \\ 0 & \displaystyle \left( \begin{array}{cc} 
0 & 1 \\ -1 & 0 \end{array} \right)_{s \times s} \end{array}\right)
\left( {\Di \6  X(Y) \0 \Di \6  Y} \right)^{-1,T}  \, .\end{array}
\pl{Formel} \ee
Any choice of the functions $X(Y)$ will lead to a Poisson tensor $P^{ij}(Y)$.
Note that one traded in the complexity of finding all  solutions to the Jacobi
identity
\rp PJacobi in favour to the existence of integral surfaces of Hamiltonian
vector fields as well as the closure of the forms $\O =P\vert_S ^{-1}$ on these
surfaces, incorporated in \rp Formel through the existence of Casimir-Darboux
coordinates $X$.

The first $k=n-s$ functions $X^A(Y)$ in \rp Formel are a (locally) complete set
of independent Casimir functions of $P$, i.e.\ of those functions which have
vanishing Poisson bracket with any other function on $N$, or, equivalently,
which are invariant under the flow of any Hamiltonian vector field $P(dF,
\cdot) \equiv \{F, \cdot\}$.

Not any choice of the remaining $s$ functions leads to different functions
$P^{ij}(Y)$. To not end up with an overcomplete parametrization of $P$,  we
have to factor out the canonical transformations on the symplectic leaves
$X^A=\mbox{const}$.  This can be most easily done by requiring that one of the
functions $X^I(Y)$ shall be the identity map.  The attainability of this gauge,
e.g.\ in the form $X^n(Y) = Y^n$, can be seen by  performing the diffeomorphism
leading to \rp Formel within two steps: Let the original coordinates in which
$P$ has  Casimir-Darboux form be $\ti X$.  Firstly we perform a canonical
transformation such that the $n$-th new coordinate $X^n$ becomes an arbitrarily
prescribed function of the old coordinates $\ti X^i$; this is always possible
as is seen by inspection of the (infinitesimal) action of a Hamiltonian vector
field on a coordinate function, given that $P$ is in Casimir-Darboux form.
After this we perform a second diffeomorphism $Y=Y(X)$ which is the identity
map in the last component ($Y^n=X^n$). Both steps together clearly provide a
completely general coordinate transformation $\ti X \to Y$.  However, the first
of them does not change the form of $P$.

Equation \re{Formel}) is a general local solution, but has the disadvantage
that it involves the inverse of matrices. For practical purposes its
applicability might therefore be restricted to lower dimensions. In the case
that $P$ has at most rank two, there is an alternative form for $P$ which
avoids taking the inverse \cite{ESI}. To derive it, let us first rewrite the
expression for the corresponding Poisson brackets in its Casimir-Darboux form:
\[ \{F,G\}={\6 F \0 \6 X^{n-1}}{\6 G\0 \6 X^n} - (F \lra G) = dX^1 \wedge ...
\wedge dX^{n-2} \wedge dF \wedge dG / d^nX  \, .  \] Under a diffeomorphism
the volume element $d^nX$ changes only by a multiplicative function so that
after a general coordinate transformation we find:
\be \{F,G\}=f dC^1 \wedge ... \wedge dC^{n-2} \wedge dF \wedge dG / d^nX  \,,
\el Formel2
where $f$ and $C^1, ... , C^{n-2}$ are arbitrary functions, the latter $n-2$
ones being obviously the  Casimir functions.

There  still is a further reason for the interest in explicit formulas such
as \rp Formel and \re{Formel2}). Although, if no coordinate system is
distinguished in $N$, all Poisson tensors obtained from the formulas
correspond to the same Poisson structure locally, they can be different from
a {\em global} point of view.  This happens precisely, when the foliations of
$N$ into symplectic leaves are topologically different. The symplectic leaves
are (at generic points) the level surfaces of the  Casimir functions.  So, by
approproiately choosing the Casimir functions within \rp Formel and
\re{Formel2}), one can systematically construct  Poisson structures, not
related to each other by a 'Poisson diffeomorphism' \cite{3Mad}.

Let us, as an application of the above formulas, find the most general Poisson
structure in a three-dimensional space $N$ which is rotation invariant with
respect to the $X^3$-axis. From \rp Formel2 we learn
\be P^{ij}=\e(ijk) f  C_{,k} \, , \el 3dim 
 where $\e(ijk)$ is the alternating symbol.  The latter is already invariant
under rotations (connected to the identity).  Now, $C$ needs to be
$SO(2)$-invariant, since its level surfaces are the integral surfaces of $P$.
Thus also $f$ has to be invariant.  So the most general $P$ which is invariant
under $SO(2)$ resp.\ $SO(1,1)$ transformations in any ($X^3=$ const)--plane
$\subset \dR^3$ is provided by Eq.\ \re{3dim}), in which the free functions $f$
and $C$ depend only on $X^3$ and $(X)^2 := (X^1)^2 \pm (X^2)^2$.

We noted already that for a linear $P$, $P^{ij}=f^{ij}{}_kX^k$, the coefficients 
$f^{ij}{}_k$ are structure constants of some Lie algebra ${\bf g}$.  
Obviously in this case the vector fields $V^i = \{ X^i, \cdot \}$
generate (co)adjoint transformations on the space $N={\bf g}^\ast$. The $V^i$
form an overcomplete basis in $TN$ and the symplectic leaves coincide with the
coadjoint orbits in ${\bf g}^\ast$. The corresponding symplectic form, 
introduced and studied by  Kirillov \cite{Kir}, Kostant and
Souriau \cite{Kos}, is determined through
\be \O(V^i,V^j)=\{X^i,X^j\}= f^{ij}{}_kX^k \, , \el Kir
and plays some role in the representation theory of Lie groups \cite{Kirbuch}.

Before  closing this section, let us consider the case that one wants to
quantize some symplectic manifold $(S, \O)$, not necessarily  diffeomorphic to
$\dR^s$; it could be any symplectic leaf of a given Poisson structure $P$ in 
$N$. Within  the framework of geometric quantization \cite{Wood} the wave
functions are sections in a Hermitian line bundle over $S$ with curvature
$\O/\hbar$.  Such a  line bundle exists, iff $\O$ is 'integral', i.e.\ iff
\be \int_\s   
\O =2\pi n \hbar \equiv n h\, ,\qquad n \in Z \, .\el quan 
for any two-surface $\s \subset S$. This is a consequence of the fact that in a
line bundle the parallel transport with respect to $\nabla$ around a closed
curve $\g$, $\exp(i\oint_\g \Th/\hbar)$, where (locally) $\Th=d^{-1}\O$, can be
equivalently expressed as $\exp( i\int_\S \O/\hbar)$, if $\6 \S = \g$.
Different choices of $\S$ then yield the necessity of \re{quan}). 
Since $\O$ is closed, this condition is empty,  
if the second fundamental group of $S$, $\Pi_2(S)$, is trivial. If $\Pi_1(S)$
is trivial, furthermore, the line bundle is unique; otherwise there arises some
arbitrariness in the quantization, which can be parametrized by the irreducible
representations of this $\Pi_1(S)$ \cite{Wood}, \cite{Bal}. 

As an example let  us regard the coadjoint orbits $S$ of $N=so(3)^\ast$.  The
coadjoint transformations are rotations about the origin and thus the
symplectic leaves $S$ are two-spheres characterized by their radius
$r=\sqrt{X^iX^i}$. Only for $r=0$ the symplectic leaf shrinks to  a point. To
evaluate \rp quan we first have to determine $\O$. This can be done most easily
by noting that $\O$ has to be rotation invariant (as rotations are generated by
Hamiltonian vector-fields) and that it has to be linear in $r$, cf.\ Eq.\
\re{Kir}).  Thus one finds
\be \O = r\sin \dh  d\dh\wedge d\ph = 
dX^3 \wedge d\ph \, , \el Oso where $r$, $\dh$, and $\ph$ are the standard
spherical coordinates.  $X^3$ and $\ph$ are  seen to be possible Darboux
coordinates for $\O$.  Here $\O$ can be rewritten as $\e_{ijk} X^i dX^j \wedge
dX^k/X^iX_i$, but, although suggestive, such an explicit formula for $\O$ in
terms of structure constants does not exist for general coadjoint orbits.
Combining \rp quan with \rp Oso we find that only the spheres of radius
$n\hbar/2$, $n \in N_0$ are quantizable symplectic manifolds.

For $r=n\hbar/2$ the  space of sections in the  line bundle over $S$ is
infinite dimensional; it is the result of prequantization. The final quantum
theory on a (quantizable) symplectic leaf $(S, \O)$ is obtained after choosing
a polarization, which is implemented as a horizontality condition on the
sections; for  a holomorphic polarization this reads $\nabla_{\bar z}
\psi(z,\bar z)=0$.  The dimension of the resulting Hilbert space is $n+1$; it
is an irreducible representation of $SU(2)$ for spin $n/2$.

Note that if instead of geometric quantization we applied a
rather algebraic approach to quantize the Poisson bracket relations
$\{X^i,X^j\}=\e(ijk)X^k$, ignoring for a moment the constraint
$X^iX^i=r^2=\mbox{const}$, we get the quantization condition $r^2=n(n+1)/4$
instead. This difference is at the heart of the ongoing discussions on the 
different spectra obtained when quantizing $2D$ Yang-Mills theory
\cite{spectra}.

The quantization condition \rp quan can be also obtained from a path integral
point of view \cite{Anton}. The action for the above $so(3)$--invariant point
particle systems is of a Wess-Zumino type, which we want to write as $L_p=\int
\Th$ where $\Th=\O$ denotes the canonical potential of the Kirillov form $\O$.
However, there is no globally well-defined $\Th$ as $\oint_S \O \neq 0$ (for $r
\neq 0$). Being interested in the path integral, we thus content ourselves 
with an action $L_p$ that is well-defined up to the addition of a multiple of
$2\pi$.  The most general $\Th=d^{-1} \O$  on $S^2$ without  poles  has the
form $r\cos \dh d\ph + \l d \ph$. The above condition on the action then leads
to $(r \pm \l) \hbar \in Z$ (point particle trajectories close to the poles)
which again yields $r=n\hbar/2$. Let me remark here in view of \re{path}) that
as an action for a two-dimensional field theory $\int \O$ obviously {\em is}
well-defined globally also for $S=S^2$.

\section{A General Action in Two Dimensions} 
\la{secaction}

As will be shown in this section, any Poisson structure $P$ on a manifold $N$
induces canonically a  topological field theory on a given two-dimensional
world sheet manifold $M$.  By means of an additional  volume form $\e$ on $M$
and a Casimir function $C^1(X)$ of $P$ one can add to this, furthermore,  a
nontrivial Hamiltonian.

On the bundle $E\equiv \L^{1,1}\left(T^\ast(M\times N)\right)$ there is a
canonical form $A$. In local coordinates $x$ and  $X$ of $M$ and $N$,
respectively, it can be written as $A= A_{\m i}dx^\m \wedge dX^i$. Let
$\Phi$ be the map from the world sheet manifold $M$ to $E$; if $\Pi_M$ denotes
the projection in the bundle $E$ on $M$, then let $\Pi_M \circ \Phi$ be 
the identity map. The topological part of the action we postulate has the
following form
\be L_{top}=\int \Phi^\ast(A + \2 i_Ai_AP) \, , \el action
where $i_A$ denotes the insertion and $\Phi^\ast$ is the pull back of the map
$\Phi$. $L_{top}$ is manifestly invariant under separate diffeomorphisms on $M$
and $N$. Note, however, that it is not invariant under (arbitrary)
diffeomorphisms on $M\times N$ due to the special form of the fiber $E$.  Since
$L_{top}$ is ($M$-)diffeomorphism invariant without the use of a (background)
metric on $M$, and since this feature holds also on the quantum level, it is
'topological'  \cite{Blau}; below it will be found to be of the Schwarz
type.

With the additional input of a ('background') volume form $\e$ and the choice
of a Casimir function $C^1$, we can extend this action with
\be L_+= \int \e \Phi^\ast C^1 \, .\el S+
In coordinates the action $L=L_{top} + L_+$ takes the form
\be L = \int_M d^2x \, \{ \e(\m\n) [A_{\m i}(x) X^i{}_{,\n}(x) + \2 P^{ij}(X(x)) 
A_{\m i}(x) A_{\n j}(x)] +  \ti \e(x) C^1(X(x)) \}  \, , \el actioncoor where
$\e(\m\n)$ denotes the alternating symbol and we defined $\ti \e$ according to
$\e(x)=\ti \e(x)d^2x$. Displaying only the $X$ coordinates, $S$ becomes 
\be L=\int_M A_i \wedge dX^i + \2 P^{ij}(X) A_i \wedge A_j +\e C^1\, , 
\el actionform2  
where we suppressed writing the pull-back.

Let us now turn to the question of local symmetries starting with the simplest
situation $P\equiv 0 \equiv C^1$. In this 
case $L=\int \Phi^\ast A$ and the most
general symmetry of the action up to a total divergence, i.e.\ up to a local
exact term, is provided by $A \to A + d\ep$.

The other extreme case is that $P$ is invertible. In this case we first try to
lift the diffeomorphisms in the $N$ space.  We noted already that $L$ is
$N$-diffeomorphism invariant. However,  when regarding symmetries of the action
as functional of the fields $X^i$ and  $A_{\m i}$, $P$ (and also $C^1$) are not
allowed to transform.  Thus we are left only with the symplectomorphisms of
$P$, i.e.\ those transformations of $X$ whose Lie derivative on $P$
vanishes.\footnote{All other $N$-diffeomorphisms, where also $P$ and $C^1$ are
transformed appropriately, correspond to a different parametrization of one and
the same action; after such a change of coordinates it is a different
functional of, e.g., $X^1(x)$.} The latter are transformations generated by
(locally) Hamiltonian vector fields and   they clearly 
leave  $A$ (by
appropriate transformations of its components) as well as $C^1$ invariant.
Since an $M$-dependent diffeomorphism in $N$ does not respect the one-one
splitting of the form $A$, these transformations do not directly transfer to
symmetries of $L$. Rather,  under a transformation generated by $\ep_i(x)
P^{ij}(X) \6 / \6X^j$ the form $A$ on $E$ picks up also a $\L^{2,0} T^\ast
(M\times N)$ part $i_Ai_{d\ep} P$. However, shifting $A$ further by $d\ep$ the
action \rp action becomes obviously invariant up to a total divergence (note
that the insertion of $i_Ai_{d\ep} P$ into $P$ is zero).

Choosing Casimir-Darboux coordinates as a (local) parametrization of $L$, one
immediately finds the general situation to be a direct  superposition of the
two cases studied above.  Thus, in a somewhat formal manner, one can write for
the symmetries of \re{action}):
\ba \d_\ep X^i &=& \{ \ep_j X^j, X^i \}_N = \ep_j P^{ji}  \equiv  i_{dX^i}
i_{d\ep} P\nn
\d_\ep A &=& d\ep + i_Ai_{d\ep}P \pl{sym1} \, , \ea
where $\ep=\ep_i(x)dX^i$. (We used the suffix $N$ for the Poisson brackets on
this space, so as to not  get confused with the Poisson brackets of the {\em
field} theory in its Hamiltonian formulation introduced below). Under these 
transformations  the action changes by $\int_M \Phi^\ast d\ep$. The appearance
of the two-zero form on the righthand side of the second equation \re{sym1}) 
is somewhat ugly.  In the
useful halfway component notation of \rp actionform2 the symmetries \rp sym1 
can be rewritten as
\ba \d_\ep X^i(x) &=&  \ep_j(x) P^{ji}(X(x)) \pl{sym2a} \\
\d_\ep A_i &=& d\ep_i(x) + P^{lm}{}_{,i} A_l \ep_m  \, .\pl{sym2b} \ea

Variation of \rp actionform2 leads to the field equations 
\ba  dX^i +  P^{ij} A_j&=&0 \pl{eom1a}\\
dA_i + \2 P^{lm}{}_{,i} A_l \wedge A_m + \e (C^1)_{,i}&=&0 \, . \pl{eom1b} \ea
A simple comparison of these first order differential equations 
with the symmetries above establishes that there will be no local degrees of 
freedom. (Eq.\ (\ref{sym2a}) containes $k$ independent local 
symmetries, if $k$ denotes the number of Casimirs, and \rp sym2b  
containes $n-k$ further ones). Thus there will be only a finite number 
of degrees of freedom and these will be of some  global nature. 

From these considerations we also see that there are no further local 
symmetries of our action. In particular, the  diffeomorphism invariance 
of \rp actionform2 for $C^1 \equiv 0$ 
has to be incorporated already within the symmetries 
(\ref{sym1}). Indeed, for any given vector field $\xi =\xi^\m(x) \6 / \6 x^\m$ 
generating diffeomorphisms on the world sheet manifold $M$, 
the (field dependent) choice $\ep := i_\xi A$ in (\ref{sym2a},
\ref{sym2b}) results in
\ba \d_{i_\xi A} X^i &\equiv& \CL_\xi X^i - i_\xi (dX^i +  P^{ij} A_j) \nn 
\d_{i_\xi A} A_i   &\equiv& \CL_\xi A_i - 
i_\xi (dA_i + \2 P^{lm}{}_{,i} A_l \wedge A_m)  \pl{diff} \, ,\ea
where $\CL_\xi$ denotes the Lie derivative along $\xi$. Obviously the 
additional terms on the righthand side of \re{diff}) vanish for any 
solution to the field equations \re{eom1a}, \ref{eom1b}) exactly for 
$C^1 \equiv 0$. 

The action $L$ is in first order form, i.e.\ it is already a Hamiltonian 
action. A Hamiltonian formulation for infinite dimensional systems 
requires appropriate boundary conditions. Therefore  we choose $M$ 
to be of the form $S^1 \times R$, at least locally.  
More general topologies of $M$ might then be obtained from an 
appropriate sewing procedure (cf.\ \cite{Wit2d}).  We parametrize 
$M$ by a 
$2\pi$ periodic coordinate $x^1$ and the 'evolution' parameter $x^0$. 
Note that this does in no way restrict $x^0$ to be 'timelike' (with respect to 
whatsoever a metric). 
As seen most directly  from \re{actioncoor}), $A_{1i}$ 
is the conjugate variable to $X^i$, i.e., with the convention $\e(01)=1$, 
\be \{ X^i(x^1), A_{1j}(y^1) \} = - \d^i_j \d(x^1-y^1) \, , \el can
and the Hamiltonian is 
\be H= \int dx^1 (\ti \e  C^1  - A_{0i}G^i) \el Ham 
with ($\6 := \6 / \6x^1$)
\be G^i\equiv \6X^i +  P^{ij} A_{1j} \approx 0 \,. \el cons
The weak equality sign '$\approx$' indicates that the $G^i(x^1)$ 
are zero only on-shell \cite{Dir}, as enforced by means of the Lagrange
multipliers  $A_{0i}$, which we may regard as arbitrary external functions on
the phase space.  By means of the Poisson brackets \rp can the constraints \rp
cons can be easily seen to generate the (one-components) of the symmetry
transformations \re{sym2a}, \ref{sym2b}). They are first class constraints,
i.e.\ they close on-shell with respect to the Poisson bracket \re{can}):
\be \{G^i,G^j\}=P^{ij}{}_{,k}G^k \d \, , \el consalg
where we suppressed writing  arguments. Let me remark that 
in order to have the first class property for \rp cons the Jacobi identity 
for $P^{ij}$, Eq.\ \re{PJacobi}), is not only sufficient but also necessary 
(whereas for $n\equiv dim N >3$, the requirement to have the  
 Hamiltonian vector fields of $P$ on $N$ to be in involution does not lead to
\re{PJacobi})).  It is now straightforward to check that the
$x^0$-evolution generated by $H$ reproduces (\ref{eom1b}) as well as the zero
components of \re{eom1a}) (whereas the one-components of Eq.\ (\ref{eom1a}) are
identical to \re{cons})).  Most of the gauge freedom has been separated to the
freedom in choosing $A_{0i}$ (axial gauge) and the constraints $G^i$ generate
the corresponding residual gauge freedom.

There is also another way to interpret the symmetries generated by the first
class constraints: Allowing for $x^0$-dependent coefficients, the $G^i$ can
generate all symmetries \re{sym2a}, \ref{sym2b}) for $X^i$ and $A_{1i}$; the
transformation of the Lagrange multiplier fields $A_{0i}$ can then be
determined --- in a closed form for a general constraint algebra --- by
requiring that the Hamiltonian action shall be invariant up to a total
derivative \cite{Hen}. The net result then coincides with \re{sym2a},
\ref{sym2b}).

Let us briefly discuss the BRS formulation of the model \re{action}).  The main
idea of the BRS technique is to enlarge the phase space by introduction of (in
our case)  fermionic degrees of freedom ('ghosts'), destroying the local
symmetry of the action, or rather turning it into a global one generated by the
nilpotent BRS charge $Q$. The 'physical' content of the theory, i.e.\ the
reduced phase space, is reobtained, when one passes to the cohomology of $Q$.
There are some advantages of the Hamiltonian BRS formulation against the
Lagrangian one: the BRS $Q$ does not depend on the chosen gauge, the BRS
transformations are canonical transformations in the enlarged phase space, and
there is the canonical symplectic measure for the definition of the path
integral. Note  that  also covariant gauge conditions may be introduced in the 
Hamiltonian formalism; one only has to trivially enlarge the phase space by 
including the Lagrange multiplier fields $A_{0i}$ together with momenta for 
them which are constrained to zero \cite{Hen}. 

In our case the BRS charge is extremely simple. Despite the appearance of 
structure functions in the constraint algebra \re{consalg}), it still 
has the minimal form 
\be Q= \eta_i G^i - \2 \eta_i \eta_j P^{ji}{}_{,k} \CP^k \, , \el BRS
where $(\eta_i,\CP^i)$ are canonically conjugate fermionic ghost variables 
associated with $G^i \approx 0$. It is straightforward to verify $\{Q,Q\}=0$
as a consequence of \re{PJacobi}). Note that, depending on the chosen 
gauge fermion $K$, there still may appear quartic ghost vertex contribution 
in the quantum action; as, e.g., it may happen in the multiplier gauge
$K=-\chi_i \CP^i$, if the gauge conditions $\chi_i(A_{1i}, X^i)$ do not commute
with $P^{ji}{}_{,k}$.

It would be desirable to extend the formulation of \rp action so as to
explicitely include the case of nontrivial fiber bundels  over $M$. $A$ can
then be  a one-one form only locally, i.e.\ only on some $U \times N$, $U 
\subset M$, because one should allow for a nontrivial fibration of $N$ over
$M$. We do not have a satisfactory answer for this problem yet. Probably there
should be some superior formulation of the model, maybe including further
fields which can be gauged out only locally, or, as in the case of the
WZW-theory,  having some three-dimensional closed term in its action. Then this
other action should be strictly invariant under the symmetry transformations
and also the transformation \rp sym1 for $A$ should be replaced by something
more perspicuous.

Some remarks might be in place here: 

Firstly, the field equations (\ref{eom1a}, \ref{eom1b}) are (strictly)
covariant under the transformations (\ref{sym2a}, \ref{sym2b}) as they stem
from a local variation of the action. It is no problem to explicitely construct
nontrivial bundles over $M$ by means of their solutions, at least if one can
integrate the infinitesimal form of the symmmetry transformations \re{sym2a}, 
\ref{sym2b}).

Secondly, as a Hamiltonian system, i.e.\ on the level of the field equations,
the completely gauged WZW theory with compact gauge group $G$, defined on the
cylinder via a Gauss decomposition, turns out to be a special case of our
theories with $N=G$ \cite{AntonPT}.  A natural question then arises: What is
the ungauged or partially gauged version of the general model (\ref{action})?

Thirdly, in the case that we choose a linear Poisson structure, $P^{ij}=
f^{ij}_{k}X^k$, we regain two-dimensional nonabelian gauge theories: In this
linear case the $N$-coordinate $X^i$ is an equally well-behaved object as
$dX^i$; $A_i$ is then commonly spanned on Lie algebra generators $T^i$,
$A=A_iT^i$, satisfying $[T^i,T^j] = f^{ij}_{k}T^k$, and similarily $X^i$ is
spanned on a dual basis.  After addition of  $\int d(A_iX^i)$ to
(\ref{actionform2}) and a partial integration, $L_{top}$ takes the standard form
\be L_{top}=\int_M X^i F_i \, , \el gaugeth2 
where $F=dA + A\wedge A$. The standard 2D Yang-Mills theory $\int tr(F \wedge
\ast F)$ for semisimple groups is obtained, in its first order form, when
choosing $C^1 \propto tr(XX)$ and setting $\e$ equal to the  metric induced
volume form used to define the Hodge dual $\ast$. Thus in this linear case, and
only there, the addition of the surface term $\int d(A_iX^i)$ provides
an action which is strictly invariant under the transformations (\ref{sym2a},
\ref{sym2b}); and certainly here we also know how to (simply) understand
nontrivial fiber bundels.

Last but not least, it might be interesting to study the limit $n \to \infty$ 
of the dimension of  target space $N$ of the model \re{action}) \cite{Peter}. 
In this way one may gain topological field theories in  higher dimensions. For 
instance, an application of such a limit to the completely gauged WZW model
was shown to result in fourdimensional selfdual gravity. 

(During the completion of this work my colleague P.\ Schaller found
the Wess Zumino type formulation of \re{action}). E.g., in the case of  the 
gauge theories \rp gaugeth2 the corresponding boundary term has the form 
$\int_B DX^i F_i$, where $D$ denotes the covariant derivative. It obviously 
reprodues \re{gaugeth2}) due to $DF\equiv 0$, if $M=\6 B$.)

\section{The Classical Solutions}  

\label{secclass}

There are basically two ways to solve the field equations \re{eom1a}, 
\ref{eom1b}). 
One can either choose gauge conditions or one can  work in appropriate 
target space coordinates, most 
referably in Casimir-Darboux coordinates of $P$ (cf.\ section \ref{sec1}). 
In any case one will have to determine the Casimir functions of $P$. So 
to start with let us suppose one knows how to cover $N$ by charts 
$U_a$ in each of which one has Casimir-Darboux coordinates 
$X^A,X^I$ of $P$. By continuity of all maps $\Phi$ from $M$ to $N$, 
any {\em global} solution $\Phi$ satisfying the 
field equations can be obtained by  an appropriate patching on $M$. 

Let the  first Casimir coordinate $X^1$ coincide with the Hamiltonian $C^1$.
The independent field equations then take the simple form ($C^1 \not\equiv 0$):
\ba dX^A&=& 0  \la{eomC1}\\ dA_1 &=& -\e 
\la{eomC2}\\ 
dA_A &=& 0, \,\,\quad \mbox{for index $A \neq 1$} \la{eomC3}\\ A_I &=& \O_{JI} dX^J
\, , \la{eomC4} \ea where  taking the pullback  is understood implicitly. In
the case $C^1\equiv 0$ one has $dA_1=0$ instead of \re{eomC2}).  Obviously the
general solution to the above equations can be obtained without any choice of a
gauge: The Casimir fields $X^A(x)$ have to be constant on $M$, but otherwise
arbitrary, whereas the $X^I(x)$ remain completely undetermined by the equations 
of motion.  Any
choice of the latter determines $A_I$ uniquely through \re{eomC4}).  For
$C^1\equiv 0$, $A_A = df_A$, finally,  with  arbitrary functions $f_A$.  For
$C^1\not\equiv 0$, there always exists a local solution to \re{eomC2}) because
$\e$ is closed, and $A_1$ is again determined only up to an exact one-form
$df_1$. (E.g.\ in coordinates on $M$ such that $\e=d^2x$: $A_1= x^0dx^1 +
df_1$).

Up to now one has not made use of the gauge freedom. As is obvious from
\re{sym2a}, \ref{sym2b}) any choice of the $X^I$ is gauge equivalent, and also
$A_A \sim A_A + dh_A$, where the $h_A$ are arbitrary functions.  Thus
{\em locally} any solution to the field equations is uniquely determined by
the values of the Casimir functions.

Additional structure evolves, if global apects are taken into account. In a
completely coordinate independent manner the field equations \rp eom1a take the
form
\be \Phi^\ast(e + i_Ai_e P)=0 \quad  \forall e \in T^\ast N \,, \el eom2a
where, Aas before, $\Phi$ denotes the map from $M$ into $F =\L^{1,1} 
T^\ast(M\times N)$, the projection $\Phi_M$ 
of which onto $M$ is trivial. Reformulating the previous local 
considerations $N$-coordinate indepently, we find  all global solutions to 
\re{eom2a}):\footnote{In the case of a nontrivial fiber bundle one merely
replaces  'map' by 'section'.}

{\bf 1)} $\Phi_N \equiv \Pi_N \circ \Phi$, where $\Pi_N$ denotes the projection
in  $F$ to $N$, may be an arbitrary map from $M$ into any symplectic leaf (or
integral surface) $S \subset N$ of $P$. All smooth deformations of this
embedding of $M$ into $S$ are gauge transformtions.

{\bf 2)} The restriction of $A$ to $T S$ , $A\vert_S$, (i.e.\ the pullback of
$A$ with respect to the embedding function of $S$ into $N$), is then uniquely
determined by:
\be \Phi^\ast[i_v (A  + \O)]  = 0  \quad  \forall v \in TS,  \el solArestr
where $\O\in\L^2T^\ast S$ denotes the inverse of $P\vert_S$.

The remaining field equations \re{eom1b}) may not that easily be rewritten in a
completely coordinate independent manner.  Still we know that in any local
Casimir-Darboux coordinate system on $N$ the only remaining equations to be
solved are \re{eomC2}, \ref{eomC3}). Also all the remaining local symmetries
can be integrated easily in this coordinate system yielding $A_A \sim A_A+
dh_A$.  In many cases this will suffice  to classify solutions globally;
examples for this shall be provided elsewhere.

In some instances it may be favorable to express \re{eomC2}, \ref{eomC3}) in a
more general coordinate system. Let  the first $k$ coordinates still be Casimir
coordinates $X^A$, but the remaining $s=n-k$ coordinates $X^\a$ be arbitrary.
Then for $i=A$ Eq.\ \re{eom1b}), with, for simplicity, $C^1\equiv0$, takes the
form\footnote{As already in \re{eom1a}, \ref{eom1b}) we suppress writing the
pullback in the  following.}
\be dA_A +  \O_{\a\b,A}dX^\a dX^\b=0 \, .\el daund
One can check further that Eq.\ \re{eom1b}) with $i=\a$ is already fulfilled by
the solutions {\bf 1} and {\bf 2} above, i.e.\ that these equations are already
a consequence of \re{eom1a}).  Locally, Eq.\ \rp daund can be integrated
easily: Up to gauge transformations one finds $A_A= \Th_{\a,A}dX^\a$, if $\Th$
again denotes a symplectic potential for $\O$.

Before turning to the quantum theory, let me briefly comment on the option of
solving the field equations by first choosing gauge conditions.  Probably the
most efficient one is $A_{0i}=0$. Locally this gauge is always attainable by
means of the symmetry transformations \re{sym2a}, \ref{sym2b}), as is most
easily seen in the Hamiltonian formulation of the theory, where the $A_{0i}$
are subject to arbitrary shifts.\footnote{In the gravity theories considered in
the following chapter the gauge $A_{0i}=0$ corresponds to a degenerate metric,
as is obvious from \re{identi}). One of the simplest choices not in conflict
with the metric nondegeneracy is $e_0{}^-=\o_0=0$, $e_0{}^+=1$, used, e.g., in
\cite{DomLC}. The consequences of the fact that the Hamiltonian symmetries 
connect degenerate with nondegenerate metric configurations is studied in
detail in Sec.\ \ref{secJT}.} For $C^1\equiv 0$, e.g.,  the field equations
state in this  gauge that all fields are $x^0$-independent and that they have
to satisfy the constraint equations \re{cons}). The general solution to the
latter is then provided by {\bf 1} and {\bf 2} above, where $M$ is replaced by
an open interval.

\section{All Quantum States}
\la{secquan}

Up to now an exact  treatment of standard quantum  field theories is beyond
human abilities and one takes recourse to approximative methods such as
perturbation theory. In our case, however, a nonperturbative treatment is
accessible.

The Hamiltonian formulation of our theory has been presented already in section
\ref{secaction}. It corresponds to a worldsheet topology $M=S^1 \times R$.  Let
us consider the wave functionals in an $X$-representation. Any quantum wave
function $\Psi$ is then a complex-valued functional of parametrized smooth
loops $\CX: S^1 \to N$ in $N$.  However, following Dirac \cite{Dir}, only such
quantum states are admissible which satisfy the quantum constraints
\be \hat G^i(x) \Psi[\CX] = \left(
\6 X^i(x) + i\hbar P^{ij}(X) {\d \0 \d X^j(x)} \right)
 \Psi[\CX] =0 \, , \el qcons resulting from  \rp cons by the replacement
$A_{1i}(x^1) \to i\hbar \d /\d X^i(x^1)$, suppressing the superscript one for
the variable $x^1$ within this section. It is decisive that the operator
ordering within the $\hat G^i(x)$ is such that taking commutators between the
quantum constraints does not produce further constraints. \rp qcons still leads
to the constraint algebra \re{consalg}).

Maybe less well-known is an additional restriction on the operator ordering
within $G$, probably present in an analoguous manner in most  diffeomorphism
invariant theories. Any gauge invariant phase space function, such as, e.g.,
the Casimir functions $C^A(X(x))$, have to be $x$-independent on-shell because
diffeomorphisms are part of the symmetry transformations (cf.\ also Eqs.\
\re{diff})). Therefore $\6 C^A(X(x))$ will be part of the field equations, and,
as such, it will result from an appropriate combination of the constraints.
Since, however, $\oint \6 C^A dx \equiv 0$, the integral over the corresponding
combination of the constraints will also vanish identically; in our case
\be 0 \equiv \oint \6  C^A dx^1 \equiv \oint {dC^A \0 dX^i}  G_i \, . \el depen
This indicates a subtle dependence of the constraints among each other.  It is
decisive to maintain the  relations \rp depen also on the quantum level, and
we have done so in \re{qcons}).  This is maybe best illustrated at the simple
example of two classical constraints $g^1=g^2=qp$, in a phase space of some
arbitrary dimension, which obviously satisfy $g^1 -g^2=0$; The operator
ordering $\hat g^1=-i\hbar q (d/dq)$,  $\hat g^2=-i\hbar (d/dq) q$ does not
produce any anomaly in the (trivial) constraint algebra, but obviously there is
{\em no} nontrivial common kernel of the constraint operators as  $\hat g^1
-\hat g^2=-i\hbar$.

Again the solution to \rp qcons is most easily found in Casimir coordinates
$(X^A,X^\a)$ of $P$. In a parallel way as we obtained solution {\bf 1} of the
previous section we find that the support of the wave functionals has to be on
such loops which lie entirely within some integral surface $S$ of $P$. Next we
have to solve
\be [{\d \0 \d X^\a(x)} - {i \0 \hbar} \O_{\a\b}(X(x)) \6 X^\b(x)]
\Psi [\CX]  =0 \, . \el qconsref1
This equation can be reinterpreted as a {\em horizontality condition} for the
complex-valued functionals on the space $\G_S$ of loops on each  $S$ (but not
on $N$, except for an invertible $P$ resulting in $S=N$):
\be (d + (i/ \hbar) \CA)\Psi =0 \, , \el qconsref2
where the $U(1)$-connection is uniquely defined via
\be \CA({\d \0 \d X^\a(x)}) = - \O_{\a\b}(X(x)) \6 X^\b(x) \, .
\el qconnection

A necessary condition for \rp qconsref2 to have nontrivial solutions is that
$\CA$ is closed. According to \rp qconnection there is a close relationship
between $\CA$, which is a connection in a $U(1)$-bundle over the loop space
$\G_S$, and the symplectic form $\O$ on the underlying space $S$. Let us make
this relationship more precise, so that finally $d\CA=0$ will be a simple
consequence of $d\O=0$.

Any two-form $\o$ on a manifold $S$ generates a one-form $\a$ on the loop space
$\G_S$ on $S$: Forms are basically the dual objects to areas of integration.
Now, any path $\g$ in $\G_S$, corresponding to a one-parameter family of loops
in $S$, obviously spans a two-dimensional surface $\s(\g)$. Thus, given $\o \in
\L^2T^\ast S$, we can uniquely define $\a$ via
\be
\int_\gamma\a = \int_{\s(\g)}\o \, ;
\el co
in this equation we assigned a number to any path $\g \in \G_S$,  which, by
duality, defines the one-form $\a$. Since, furthermore, any closed path $\g$
corresponds to a closed two-surface $\s(\g)$, $\alpha$ is closed, iff $\o$ is
closed.

Of course, not every one form on $\Gamma_S$ can be described in this way.  In
our case, however, the one form ${\cal A}$ on $\Gamma_S$, is indeed generated
by $\O$.  To prove this let us choose a path $\gamma \in \Gamma_S$ parametrized
by a parameter $\tau \in [0,1]$. Any point in $\gamma$ corresponds to a loop
$\CX$. Thus $\gamma$ induces a map $S^1\times [0,1] \to N: (x,\tau)\to
X(x,\tau)$, which precisely corresponds to a parametrization of $\s(\g)$
introduced above. Denote by $\dot \CX \in T\G_S$ the tangent vector to $\g$:
$\dot \CX=\oint dx \dot X^\a(x,\t) (\d / \d X^\a(x))$, where the dot denotes
the drivative with respect to $\t$. Then, as an obvious consequence of
\re{qconnection}),
\begin{eqnarray} \lefteqn{\int_\g {\cal A} \,\, = \,\,
\int_0^1 {\cal A}(\dot {\cal
X})d\t \,\,=} \nn  &= & - \int_0^1\oint_0^{2\pi}
\dot X^\a(x,\t) \O_{\a\b}(X(x,\t)) \6 X^\b(x,\t) \, dxd\t \, = \,
 \int_{\s(\g)} \O \, . \label{rela} \end{eqnarray}

Thus \rp qcons can be integrated locally to yield $\Psi = \exp
\left[(i/\hbar) \int \CA \right] \Psi_0$ for any initial value
$\Psi_0$. The integrability extends to a global one, if $\G_S$ is simply
connected, i.e.\ if $\Pi_2(S)$ is trivial. For the case that $\Pi_1(\G_S)
\equiv \Pi_2(S) \neq 1$, however, the one-valuedness of a nontrivial  $\Psi$ is
given, if and only if $\CA$ is integral, i.e.\ iff
\be \oint_\g \CA = nh \, , \quad n \in Z    \el aquan
for any (noncontractible)  closed loop $\gamma$ representing an element of
$\Pi_1(\Gamma_{S})$. Due to \rp rela this is equivalent to an integrality
condition for $\O$:
\be \oint_\s \O = nh \, , \quad n \in Z    \el oquan
for any (noncontractible)  closed two-surface $\s$ representing an element of
$\Pi_2(S)$.

Let us denote the space of symplectic leaves by $\CS$. As this is the space $N$
modulo the flow of the Hamiltonian vector fields, $\CS$ in general is no more a
smooth manifold. Nevertheless, at least when ignoring the symplectic leaves of
less than maximal dimension $s$, any atlas $(X^A, X^\a)$, where again the $X^A$
are local Casimir coordinates, can be used to define an atlas of $\CS$ with
local coordinates $X^A$. ($X^A=$ const characterizes then all of the orbit,
even if it leaves the chart $(X^A, X^\a)$).   Now, if some symplectic leaves $S
\in \CS$ have a nontrivial second fundamental group, the integrality condition
\rp oquan may yield a restriction of the support of $\Psi$ to loops on
a (possibly discrete) subset of $\cal S$, which we name $\tilde \CS$.  For $S
\in \tilde \CS$, $\Psi$ is determined up tp a multiplicative constant on any
connected component of $\Gamma_S$.  As the space of connected components of
$\Gamma_S$ is in one to one correspondence with the first homotopy group of
$S$, we may identify physical states with complex valued functions on $\cal I$
defined via
\be
 {\cal I}=\bigcup_{S\in \tilde \CS} \Pi_1(S) \, \quad \tilde \CS =\{ S\in \CS :
\O \, \, \hbox{integral}\} \,\,.
\el sphst

If $\Pi_2(S) = 1 \, \forall S \in \CS$, then $\tilde \CS = \CS$. Using
the atlas  for (the generic parts of) $\CS$ introduced above in this case,
in local coordinates the
physical  wave functions take the form
\be \Psi[\CX]= \Psi_0(X^A,n_A) \exp \left[(i/\hbar) \int_D \O \right] \quad
, \qquad \6 D = \mbox{Image} \CX \, .\el wavfunc
Here $n_A$ denotes a discrete index labelling the
elements of $\Pi_1(S)$ and, as a consequence of \re{oquan}),
the choice of the disk $D \subset S$, whose boundary is the considered loop
$\CX$ of the functional, is arbitrary. In the case that $\Pi_2(S)$ is nontrivial
for some $S \in \CS$, the wave functions  may again be represented
by \re{wavfunc}) where, however,  the range of the coordinates $X^A$
has to be restricted such that \rp oquan is  fulfilled; certainly this can have
the effect of partially replacing some of these coordinates by discrete labels.

Note that the phase in \re{wavfunc}), which is not invariant under
general classical $P$-morphisms, basically coincides with the action of
the point particle system on the symplectic leaf studied in Sec.\ 2.1.
The reason for obtaining the same integrality condition \re{qcons})
can be understood by observing that in any case we want the phase factor
to be well-defined; in the case of the point particle system, so as to
have a well defined path integral, and in the case of the topological field
theory, so as to obtain a smooth wave-functional. Note also the difference
between the two quantum theories: The point particle system is defined
inherently on a symplectic leaf and, e.g.,  for an $su(2)^\ast$-orbit
of radius  $n \hbar/2$, the Hilbert space is of dimension $n+1$.
The wave functions of the field theory, on the other hand, basically reduce to
functions on the space $\tilde \CS$ of (quantizable) symplectic leaves; for
$N=su(2)^\ast$, the number of independent states on any of the  (quantizable)
symplectic leaves $S\sim S^2$ is just one in this case, since $\Pi_1(S^2)=1$.

The Hamiltonian (\ref{Ham}), being constant on each of the symplectic leaves,
defines a function on $\cal I$ and thus becomes a multiplicative operator upon
quantization. Considerations on constructing a measure on $\CI$ shall be taken
up elsewhere, or will be discussed for some of the  models considered
below (cf.\ Sec.\ 3.1).

In the example of nonabelian gauge theories (cf.\ also \cite{Ama}) $N$ is the
dual space ${\bf g}^*$ of the Lie algebra ${\bf g}$ of the gauge group.   The
symplectic leaves are the coadjoint orbits equipped with the Kirillov
symplectic form $\O$.  The integrality condition \re{qcons}) selects those
symplectic leaves \cite{Kir} which are characterized, in the case of a
compact semisimple ${\bf g}$, by values of the Casimir constants lying in the
weight lattice; the quantization of these symplectic leaves yields the unitary
irreducible representations of ${\bf g}$. This observation establishes a
connection between our representation and the $A_{1i}$-connection
representation of quantum mechanics for nonabelian gauge theories on a
cylinder \cite{YM}. Details, including a comparison of the spectra (cf.\ also
\cite{spectra}), shall be provided elsewhere.

In the noncompact case of, e.g.,  ${\bf g}=so(2,1)$, all the level surfaces of
the Casimir $C=(X^1)^2 - (X^2)^2 + (X^3)^2$ have trivial $\Pi_2$. Thus
\re{qcons}) is empty in this case. For $C>0$ we have $\Pi_1 \sim Z$,
furthermore. To display the wave functions we again choose an atlas in $N$
(inducing an atlas in $\CS$): Let us choose $(X^1,X^3,C)$ for $X^2>0$ as one of
our charts ($C=$ const $<0$ corresponds to two symplectic leaves, which may be
distinguished by the sign of $X^2$). We then have $\Psi_0=\Psi_0(C,n)$, where
$n \in {\bf N}$, and $\Psi_0(C<0, n \neq 1)=0$ since $\Pi_1(S_{C<0})$
is trivial. If one also wants to display $\int_D \O$ explicitely, one needs
further charts on $S$ ($\O_{\a\b}$ might not be well-defined, e.g., in
coordinates $X^1,X^3$ on all of the chart $X^2>0$).  In the present case of
$so(2,1)^\ast$ on any of the symplectic leaves except for the origin
 $\O$ can be  completely
described by \be \O=\pm dX^3 \wedge dX^\pm/X^\pm \, , \el Omega where
\be X^\pm = {1 \0 \sqrt{2}}(X^1 \pm X^2) \, , \el light
whereas $\O=0$ at the singular orbit $X=0$.

Having found the kernel of the quantum constraints corresponds to dividing out
all local symmetries \re{sym2a}, \ref{sym2b}) connected to the identity. Thus
the Yang-Mills theories on a cylinder described in this way are the ones
for the {\em universal covering group} of the chosen Lie algebra ${\bf g}$. The
transition to Yang-Mills theories on a cylinder for not simply connected
gauge groups calls for further steps; one has to require an appropriate
transformation of the wave functions under large gauge transformations, which
may further exclude some wave functions. For
the case of $SO(2,1)$ this has been studied in some detail in \cite{9402};
the effect is quite drastic: whereas the Yang-Mills Hamiltonian $C$ obviously
has a continuous spectrum in the case of $\widetilde{SO}(2,1)$ (=univ.\ cov.\
of $SO(2,1)$), the spectrum becomes discrete for $C<0$ in the $SO(2,1)$ theory.

\chapter{Models of Gravity in 1+1 Dimensions}

\label{cha3}

\section{A Universal Gravity Action and Remarks on its Quantization}
\label{secgravaction}

Let us construct  actions for gravitational theories from the action
$L$ introduced in Sec.\ \ref{secaction}. The starting point shall be
the identification \re{identi}), where $e_a$ and $\o$ are the zweibein and
spin connection of the gravity theory, respectively.
This requires a  target space $N$ of a minimal dimension three.

Our conventions
concerning the gravity theories shall be summarized as follows:
The metric $g$ is obtained through $g=\h^{ab} e_a \otimes e_b$, where $\h$ is
the frame metric with signature $(1, \pm 1)$. In the Minkowski case
negative lengths shall be
interpreted as spacelike distances. Since the structure
group of the frame bundle is abelian in two dimensions, the spin connection
one-form introduced above has no frame indices and the curvature is just $d\o$.
Contact with formulas used in higher dimensional gravity can be established via
$\, \o_{ab} =
\e_{ab}  \,\o$, where $\e_{ab}$ are the covariant components of the $\e$-tensor
$\e=e^1 \wedge e^2$ ($\Rightarrow \e_{12}=1$).\footnote{Note that in this
chapter $\e$ denotes
the metric induced volume form and, in contrast to the previous chapter,
it is dynamical now.}
So one finds, e.g., that the Hodge dual of the curvature
two-form, $\ast d\o$, equals the half of the Ricci scalar $R$.
For reasons of completeness we will discuss gravity theories for both
signatures of the metric in
this section, that is Minkowski as well as Euclidean gravity. In the
following section, however, we will deal with the
 Minkowski type theories only.  In this case then it
will prove useful to introduce light cone coordinates
\be e^\pm = {1 \0 \sqrt{2}} (e^1 \pm e^2) \el lightcone
in the frame bundle, which lead to an off-diagonal frame metric
$\eta_{+-}=1$ as well as to $\e^{+-}=1$.

The action of a gravity theory has to be invariant against
($M$)--diffeo\-morphisms  and frame-rotations.  The first condition leads to $C^1
\equiv 0$, i.e.\ to an on-shell vanishing Hamiltonian (cf.\ Eq.\
\re{Ham})). The
second condition has been solved in generality in the paragraph of Eq.\
\re{3dim}).  Thus the gravity action we propose has the form of $L_{top}$,
i.e.\ of Eq.\ \re{actionform2}) with $C^1 \equiv 0$, where the Poisson structure
$P$ is defined through \rp 3dim by specifying $f$ and  $C$ as functions of
\be (X)^2 \equiv X^aX_a \equiv (X^1)^2 \pm (X^2)^2  \el X2
and $X^3$.

Let us find the subclass of this family of actions which leads to a
torsion-free gravity theory. This is  equivalent  to the search for those
actions $L$  with $X^a$-dependence
$\int X^aDe_a$, where $De^a \equiv de^a + \e^a{}_b \o e^b$ is the torsion
two-form. (One could replace the $X^a$ also by strictly monotonic functions
of them, but this does not change the resulting gravity theory).
The occurence of such a
term in the Lagrangian requires $f=1 / (2 \6 C / \6 (X)^2)$.
To have it be the only $X^a$-dependent term in $L$, we find that $L$ has to
be of the form
\be L=\int_M  X_a De^a + X^3 d \o - V \e \, , \el gravaction
where $V$ is an arbitrary potential of $X^3$. This
Lagrangian was first
proposed  in \cite{Banks}.

If we again release the torsion zero condition,
$V$ in \rp gravaction may be chosen as an arbitrary function of $X^3$
and $(X)^2$ in order to correspond to a rotation invariant Poisson structure;
this can be checked by verifying \re{PJacobi}) directly, whereas the
determination of the functions $f$ and $C$ of \re{3dim}) yielding this
Poisson structure appears to be cumbersome.

All of what follows
 will be based on the action  \re{gravaction}) with
\be V=v(X^3) + {\a \over 2} (X)^2 \, , \la{V} \ee
where $v$ is some arbitrary function. The second term in this potential
allows to add a torsion squared term to the most general torsionless action
within our framework (choosing  a three dimensional target space).
The Poisson structure yielding \rp V results from the choice
\be C = (X)^2 \exp(\a X^3) +
2 \int^{X^3}_0 \! v(y)  \exp(\a y) dy  \la{q} \ee
for the Casimir function $C$ and $\exp(-\a X^3) /2$ for the integrating factor
$f$. It might be worthwhile to investigate in how far the results found below
 may be generalized to the case of an arbitrary $((X)^2, X^3)$-dependence of
$f$ and $C$.

From \re{gravaction}, \ref{V}) one can easily regain other well-known
theories of two-dimensional gravity. For instance, integrating out the Lagrange
multiplier fields $X^a$, the choice $V=\L X^3$ yields the Jackiw-Teitelboim
model (\ref{JT}). The most general quadratic potential
$V^{KV}=\mp  (X^3)^2 -\L \mp \a  (X)^2/2$
leads upon elimination of the $X$-coordinates to the Katanaev-Volovich model
\re{KV}) (use $\ast \e =\pm 1$).\footnote{Let us remind the
reader that the elimination of fields
through their {\em own} equations of motion within an action is always
possible,   at least on the classical level, as is clear from the variational
principle. For the case that the eliminated fields appeared at most quadratic
within the original Lagrangian, the same result  is obtained when integrating
out these fields within the path integral.}
In a similar way one obtains the action $L^{R^2}$ for $R^2$-gravity
\re{R2}) from the potential $V^{R^2}= (X^3)^2 -\L$. It 
coincides with \re{KV}) for Minkowski signature, if the
torsion squared term is replaced by a torsion zero condition
(limit $\a \to 0$). 

Also the model \re{string}) can be included: It 
may be obtained from  \re{gravaction}) with $V=\L$, if one replaces   
\re{identi}) by $A_a=\sqrt{X^3} e_a$ and identifies $X^3$ with $\exp(-2\Phi)$ 
\cite{Ver}.\footnote{It turns out that it is {\em always} possible to insure 
$X^3>0$ by means of gauge transformations.}  
Promoting $\L$ to a dynamical field and adding $A_4 d\L$ to 
the latter action, furthermore, this action 
can be brought even into the 
group theoretical  form \rp gaugeth2 for the centrally extended Poincar\'e 
group  \cite{Jac}. Allowing for a dynamical constant $\L$, one also may 
obtain a to \re{string}) on-shell equivalent action from the (more 
straightforward) identification \re{identi}) and $X^3 \sim \Phi$: 
Choose $V=\exp(\l X^3)$ and add $A_4d\l$ to \re{gravaction}).

But not only purly two-dimensional models may be incorporated.
The choice $V=1/2(X^3)^2$, e.g., yield solutions which are
precisely of the Schwarz-schild form, disregarding the rotation
invariant part $r^2d(\cos \th)d\varphi$; moreover, the Casimir
constant may be chosen so as to basically  coincide  (for one of its signs)
with the standard Schwarz-schild mass. In this way it is possible
to come to a quantum description of 
the Schwarz-schild soltution (i.e.\ of the {\em
spherically symmetric} solutions to the four dimensional
Einstein vacuum equations) on topology $S^2 \times S^1 \times
\dR$.

Let us conclude this section by discussing the quantization of the gravity
models \re{gravaction}, \ref{V}). We analyzed already the prototypes $V=\L X^3$
(yielding the action of an $SU(2)$ resp.\ $\widetilde{SO}(2,1)$ nonabelian
gauge theory). The generalization is quite straightforward.  Our first task is
to determine the topology of the symplectic leaves, in particular their first
and second homotopy groups. As we are in a three dimensional target space $N$
the symplectic leaves $S$ are either two- or zero-dimensional. The latter
occurs whereever $P=0$, i.e.\ at points
\be X^1=X^2=0 \, , \,\, X^3=B_{crit}=\mbox{const,} \,\, \mbox{such that} \,
v(B_{crit})=0 \,,\el crit
which we will call 'critical' or 'singular' in the following;
the corresponding value of the
Casimir constant \rp q is $C_{crit}= 2 \int^{B_{crit}}_0 \! v(y)
\exp(\a y) dy$. At 'noncritical' values of $C$ the connected
components of the level surfaces $N_C$ coincide with the symplectic leaves.

To determine the topology of the level surfaces $N_C$ it is convenient to
consider the function
\begin{eqnarray}
\lefteqn{h(X^3):=\exp(2\a X^3) \, (X)^2(X^3,C) \equiv } \nn
&&\equiv \, \exp(\a X^3) \, [C - 2 \int_0^{X^3}v(y)\exp(\a y)dy] \label{h} \ea
obtained from inverting the Casimir function \re{q}) for constant $C$.
It is only the number and kind of zeros of $h$ which determines the topology
of $N_C$, where the zeros of $h$ are nonsimple only at critical points
$X^3=B_{crit}$.

 For Euclidean signature, resulting from the
plus sign  in \re{X2}), one only has to rotate the positive parts of the curve
$\exp(-2\a X^3) \, h(X^3)$ around the $X^3$-axis to obtain the level surfaces
$N_C$ of the Casimir function.  Obviously a positive part of $h$ between
two successive simple zeros at $X^3 =B_{1,2}$ yields a symplectic leaf
isomorphic to a two-sphere. Such a leaf is subject to the integrality condition
\re{quan}), which here takes the simple form
\be \oint \O = \int dX^3 \wedge d\ph = 2\pi [B_2 - B_1] = nh \, \ee
as can be seen by expressing $P$ in coordinates $(C,\ph, X^3)$, with $\tan \ph
= (X^2/X^1)$. A positive (part of) $h$ with no (one simple) zero results in a
cylindrical (planar) symplectic leaf; the integrality condition is trivial
then. Changes of the topology of $N_C$ (along the choice of $C$)  occur only at
sliding intersections of $h$ with the $X^3$ axis, i.e.\ at critical values of
$C$.

For Minkowski signature the transition from  $h$ to $N_C$  is a bit more
cumbersome. The result is, however, quite simple: If $h$ contains no points
$(X)^2=0$, $N_C$ consists of two disconnected 'planes'; if $h$ contains $l$
points of (nonsliding) intersections with the $X^3$ axis, it has $l-1$
fundamental noncontractible loops. The second fundamental group is, moreover,
always trivial. For the critical values $C=C_{crit}$ (sliding intersections) we
again have fixed points of the Hamiltonian vector fields at $(0,0,B_{crit})$;
the set $\CS$ of integral surfaces becomes  nonHausdorff there, as can be seen
already for the $so(2,1)$ prototype (in the space of coadjoint orbits, which
here coincides with the space of Lorentz orbits in $\dR^3$,  the origin and the
positive and negative light cones have no disjoint neighborhoods).

Let us discuss  $R^2$-gravitiy, $V^{R^2}=(X^3)^2 -\L$, to
more detail. An analysis of the function $h(X^3)=C +(2/3)(X^3)^3 -2\L X^3$
yields five  qualitatively different cases depending on the parameters $C$ and
$\Lambda$ (see Fig.\ 1):

\halign{{\bf R#}:\quad&#\hfill\cr
      1 &one single zero of $h$ and thus also of $(X)^2$ at $X^3=B$ \cr
2 &one triple zero
at 0            \cr
3 &one
single zero at $B_1$ and one double zero at $+\sqrt\Lambda$ \cr
4 &one double
zero at $-\sqrt\Lambda$ and one single zero at $B_3$   \cr
5 &three single zeros at $B_1$, $B_2$, and $B_3$,\cr} where
$B_1<-\sqrt\Lambda<B_2<+\sqrt\Lambda<B_3$ and $-\infty<B<+\infty$.

Obviously $B_{crit}=\pm \L$ and the curve  along {\bf R2,3,4} in Fig.\ 1
corresponds to the critical values $C_{crit}=C_{<(>)}
\equiv
\pm(4/3)\Lambda^{(3/2)}$ of $C$.  For $\L < 0$ the quantum
theory is paritcularily simple: For both signatures the symplectic leaves are
isomorphic to planes, the spectrum of $C$ is $\dR$, and, up to the phase
factor, the wave functions \rp wavfunc are functions of one argument $C$.  For
$\L >0$ and $C
\in (-\infty,C_<) \cup (C_>,\infty)$ the resulting surfaces are again manifolds
with trivial topology.  However, for $C \in (C_<,C_>)$  and Euclidean signature
we get two disconnected surfaces of the topology of a plane and a sphere,
respectively.  Thus  the continuous spectrum $C \in {\bf R}$ has a twofold
degeneracy for some specific values of $C$ within this range $(C_<,C_>)$.  For
Minkowskian signature and $C\in (C_<,C_>)$ the level surfaces $N_C$ are
connected  and of trivial second homotopy; however, there are two fundamental
noncontractible loops, the winding numbers of which give rise to  a quantum
number $n_C \in {\bf Z}$ within the wave functions \re{wavfunc}).

Concerning the question of the inner product, let us remark here
that on large parts of the phase spaces of any of the models
\re{gravaction}) with \re{V}) and Minkowski signature, the  variable
conjugate to $C$ can be written as
\be p=-{1 \0 2} \oint \exp(-\a X^3) {e_1{}^-\0 X^-}
dx^1 \approx -{1 \0 2}
\oint \exp(-\a X^3) {e_1{}^+\0X^+}      dx^1.  \la{P2} \ee
Pulling through the phase factor of \re{wavfunc}), which in local
target space coordinates takes the form
\be \exp \left(- {i \0 \hbar} \oint\ln | X^- |\6 X^3 dx^1 \right)
\sim \exp \left({i \0 \hbar} \oint \ln | X^+ |\6 X^3 dx^1 \right) \, , \ee
the Dirac observable $p$ acts via $(\hbar/i) (d/dC)$ on $\Psi_0$.
Requiring that it  will become a hermitian operator severely
restricts the measure of the inner product, but, in  the case
that $\Psi_0$ depends also nontrivially on quantum numbers $n_C$
and/or that the level surfaces $N_C$ have several disconnected parts,
this does not determine the inner product entirely.

In the case of Minkowskian $R^2$-gravity with $\L<0$ there are no such
quantum numbers and the measure becomes the ordinary Lebesgue measure in $C$
by the above prescription; thus this theory reduces entirely to the one
of an ordinary point particle system on the line.

For Minkowskian $R^2$-gravity with $\L>0$ it is not so clear how  to
determine the inner product between states of different winding numbers $n_C$.
In this context it seems appropriate to mention that the assignment of
winding numbers at the critical values of $C$ is somewhat ambiguous:
On the one hand the critical points $(0,0,\pm \sqrt{\L})$ constitute integral
surfaces by themselves and loops in the support of $\Psi$ may not pass these
points. On the other hand the critical points and the rest of the orbit(s)
at this value of $C=C_{crit}$ do not have disjoint neighborhoods in $\CS$; so
continuous functions $\Psi_0$ identify them. Let us further note  that
a Faddeev-Popov kind of prescription for the inner product always will assign
measure zero to the loops on singular points; this seems questionable at least
in the case $v \equiv 0$, where the Poisson structure vanishes on
all of the $X^3$-axis. A clarification of these points seems desirable.

Further remarks and investigations concerning  the quantization of the gravity
theories \re{gravaction}, \ref{V}) will be made in the course of this chapter
after having explored the classical solutions into some detail.

\section{General Solution to the Field Equations and All Extremals}

As pointed out already in Sec.\ \ref{secclass}, given the present
$\s$-model-like formulation of the gravity models  the most straightforward way
to determine the classical solutions is obtained by an appropriate choice of
target space coordinates. In the present case let us  choose coordinates
\be (C,X^\pm,X^3) \, , \el targetcoo
where $C$ is the Casimir coordinate \re{q}); the transition from the original
$X^i$ to these coordinates is well-defined for $X^\pm \neq 0$. In
coordinates \rp targetcoo the symplectic form is given by \rp Omega on any of
the level surfaces $C=$ const, except at the critical points described
already in the previous section where it certainly vanishes. The usage of the
coordinates \rp targetcoo is more favorable than the one of Casimir-Darboux
coordinates since it allows to cover larger parts of $N$ and the charts with
$X^+ \neq 0$ and $X^- \neq 0$ have a common overlap, constituting an atlas of
$N$ except for  the line $X^+=X^-=0$.

In target space coordinates $(C,X^+,X^3)$ the field equations take the
simple form
\be dC=0\, ,\,  \, dA_C=0\, ,\,  \, A_+ = {dX^3 \0 X^+} \, , \, \,
A_3=- {dX^+ \0 X^+} \el eom
as  collected from \re{eomC1}, \ref{solArestr}, \ref{daund}) and
\re{Omega}).
The local solution is obvious: $C=$ const, $A_C=df$, $A_{+,3}$ as above,
and $f$, $X^+$, and $X^3$ are arbitrary functions, subject, however,
 to arbitrary gauge transformations (respecting the metric nondegeneracy).
Via \rp identi we find the relation of the  zweibein and the spin connection
to the transformed connection:
\ba e^- \equiv e_+ &=& A_+ + 2 \exp(\a X^3) X^- A_C \nn
    e^+ \equiv e_- &=& 2 \exp(\a X^3) X^+ A_C \nn
   \o &=& A_3 + 2 \exp(\a X^3) V A_C\, .
\label{relation} \ea
From this we now read off the metric
\be g \equiv 2e^+e^- = 4 \exp(\a X^3) dX^3 df + 4 h(X^3,C) df df \,
,\el metric1
where $h$ is the function defined in \rp h and
we suppressed introducing a symbol for the symmetrized tensor product.

For the torsion-free case $\a = 0$ we may now choose the gauge
\be X^3= x^0 \, , \,\, X^+ = 1 \, , \,\, f = x^1 /2 \el gaugeA
respecting $X^+ \neq 0$ and $\det g \neq 0$. The metric \rp metric1
then takes the form
\be g_{\alpha\beta}= \pmatrix{0&1\cr 1&h(x^0)} \, .\el metric
The rest of the fields is determined trivially through \re{eom},
\ref{relation}) and, using \re{h}),  $X^- = (1/2) h(x^0)$.

In the case $\a \neq 0$ it is the gauge
\be X^3= {1 \0 \a}
\ln(\a x^0) \, , \,\, X^+ = 1 \, , \,\, f = x^1 /2 \el gaugeB
with $\a x^0 \in \dR_+$ which again allows to write the metric in the
form \re{metric}); here we have set somewhat sloppily
\be h(X^3(x^0)) =: h(x^0) \, ,\el h2
noting that the degree of the zeros of $h$ is not changed by this
substitution. The rest of the fields can  again be read off directly
from \re{eom}, \ref{relation}) and \re{h}).

The result above is obtained in an almost equally straightforward  manner
when one starts from the field equations in the original coordinates $X^i$
and uses Polyakov's light-cone gauge in
the form $e_0{}^- =0, \, e_1{}^- = e_0{}^+=1$; the integration of the  field
equations turns out to be trivial in this gauge and the above found
representatives result from a fixation of the residual local gauge freedom.
Similarily appropriate is the use of the axial gauge
$e_0{}^-=\o_0=0$, $e_0{}^+=1$ \cite{DomLC}.

The analogous shape of the solutions in target space coordinates
$(C, X^-,X^3)$, valid on any patch with $X^- \neq 0$, is obtained most
easily by applying the transformation
\begin{eqnarray}
             e^+    &\longleftrightarrow& e^- \nn
             \omega &\longleftrightarrow& -\omega \nn
             X^+  &\longleftrightarrow& -X^- \nn
             X^3 &\longleftrightarrow& X^3\quad,  \label{5}
\end{eqnarray}
to the above, since \rp 5 reverses only
the sign of the action integral \re{gravaction}) and
therefore does not affect the equations of motion.
Clearly the form of the metric remains that of \rp metric under the
transformation \re{5}). Although \rp 5 is some 'cockscrewed' Lorentz
transformation  (resulting from the parametrization $\o_{ab}=\e_{ab}$ and
the fact that $\e_{ab}$ is a 'pseudo-tensor'), it may be interpreted also
as an active symmetry transformation mapping different patches of the space
time manifold onto each other; this fact will be useful in order to
extend our local solutions to global ones.

Note also that in both charts $X^\pm \neq 0$ the whole solution is
independent of $x^1$. Thus there is a Killing field,
$\partial\over\partial x^1$, generating shifts in the $x^1$-direction.

Simple zeros of $X^a$ (both components) will occur when gluing together the
above solutions. So, although possible in an analogous manner,
it is not necessary to  construct such local solutions here.
For zeros of $X^a$ of a higher degree the $a$-components of the field
equations \rp eom1a
\be dX^a + \e^a{}_b (X^b \o - V e^b) =0   \ee
together with the fact that the $e^a$ are linear independent (metric
non\-de\-generacy) show that the considered point is a singular point of $P$
(cf.\ Eq.\ \re{crit})). Since
such points in the target space constitute an integral surface $S$
by themselves and since the image of the worldsheet lies entirely in a
symplectic leaf $S$ (cf.\ Sec.\ \re{secclass}) solutions in the neighborhood
of zeros of $X^a$ of degree at least two have the form
\be              X^a =0 \, ,\,\,
              X^3 = B_{crit}=\hbox{const.}
\el momdeSitter
They are, furthermore, solutions of vanishing torsion and constant curvature,
\be     De^a =0 \, ,\,\,  d\omega = v'(B_{crit}) \e \el deSitter
describing arbitrary  deSitter space-time manifolds. The metric for such
a solution can also be brought into the form \rp metric  with
$h(x^0)= c - v'(B_{crit}) (x^0)^2$ where $c$ is some meaningful constant
of integration. This in turn determines the zweibein and spin connection
up to Lorentz transformations.

To gain some feeling for gravitational solutions as well as to construct and
analyze Penrose diagrams, it is standard to consider the movement of
 point particles within the space-time $M$ determined by the (unperturbed)
gravity solutions. For this purpose we couple $L_p[x(\t)]=m \int
\dot x^\m(\t) x^\n(\t) g_{\m\n}(x(\t)) \, d\t$ to our gravitiy action
\ref{gravaction}. Variation for $x(\t)$ leads to the standard equation for
extremals
\be \ddot x^\m  +  \G^\m{}_{\n\r} \dot x^\n \dot x^\r =0 \el extrem1
 where
$\G_{\m\n\r} \equiv (g_{\m\n,\r} + g_{\r\m,\n} - g_{\n\r,\m})/2 \,$ are
the Christoffel symbols.
Disregarding the backreaction of the point particle on the metric ($m
\leadsto 0$), we may combine this with the previously obtained solution for
the metric, Eq.\ (\ref{metric}), to get ($ h'\equiv dh(x^0)/ dx^0$)
\be \ddot x^0=-h'\dot x^1 (\dot x^0+ {1\over2}h \dot x^1)
   \quad, \qquad
             \ddot x^1 =   {1\over2}h'(\dot x^1)^2 \, .
\el extrem2

Eq.\ \rp extrem1 yields those curves which maximize the arclength $s$ for
nonnull extremals; the parameter $\t$ is, up to linear transformations, the
arclength itself \cite{SexUrb}. (Obviously \rp extrem1 is not
form-invariant under reparametrizations,
as is not  $L_p$).  In the torsion-free case these extremals
coincide with autoparallels, satisfying,  in the parametrization of
\re{extrem1}),  $\nabla_{\dot x} \dot x = 0$. In the case of nonvanishing
torsion the extremals are autoparallel only with respect to the
Christoffel connection $\G_{\m\n\r} =\o_{\m(\n\r)}$, where the
brackets indicate symmetrization.

Since $\G$ is a metrical connection, any null-line $0=g(\dot x, \dot x)$
is a solution to \re{extrem1}, \ref{extrem2}). In the charts of \rp metric
these {\it null extremals} are:
\ba   x^1&=&\hbox{const} \, , \pl{null1} \\
{dx^1\over dx^0}&=&-{2\over h} \qquad \forall x^0 \,\,  \mbox{with} \,\,h(x^0) \neq 0
  \pl{null2a} \\
 x^0&=&\hbox{const} \, ,\quad \mbox{if} \,\, h(x^0)=0  \, .\pl{null2b} \ea
Plugging these solutions back into (\ref{extrem2}), we can decide under which
conditions they are complete (with respect to the affine parameter
of (\ref{extrem2})). For \rp null1 and \rp null2a the affine parameter
is determined by $x^0= a \t+b$, $(a,b) = \hbox{const}$, thus these null extremals
 are complete, iff the coordinate $x^0$ extends to infinity into both
directions of the charts in  which the metric takes the form \re{metric}). For
the models under study, this is the case for $\a =0$, whereas in the case of
nonvanishing torsion, $\a \neq 0$, the extremals are incomplete at $x^0=0$.
Note, however, that in the latter case this line is a true singularity as
curvature and torsion blow up there.
The extremals (\ref{null2b}), on the other hand,  are complete for $h'=0$
(multiple zero  of the
function $h$),   but for $h'\neq 0$ we find $x^1= -2 (\ln \! \! \mid \!
\t + a \! \mid) / h' + b$, where $(a,b) = \hbox{const}$, so that they are
incomplete at $\mbox{sgn}(h'(x^0)) \,  x^1 \to +\infty$.

It will turn out that in our case the knowledge of the
null extremals already suffices to find all Penrose diagrams.
Nevertheless, also all other extremals can be found: Since
$\partial\over\partial x^1$ is a Killing field there is a
constant of motion, $g({\partial\over\partial x^1},\dot x)
\equiv \dot x^0+h\dot x^1=
\hbox{const}$; this holds in the case of nonvanishing torsion, too, where
one can use the Christoffel connection for the proof
(cf., e.g., \cite{Thi}).
Furthermore, we know that for nonnull extremals we may choose the length as
affine paramter so that we obtain:
$$ (dx^0+hdx^1)=\hbox{const }\, ds = \hbox{const} \, \sqrt{|2dx^0dx^1+h(dx^1)^2|} \,.
 $$
The resulting  quadratic equation has the solutions
\ba {dx^1\over
dx^0}&=&{-1\pm\sqrt{\displaystyle{c\over c-h}}\over h} \quad,\quad c=\mbox{const}
\pl{exsol1} \\
 x^0&=&\hbox{const},\quad \mbox{if} \,\, h'(x^0)=0 \, . \pl{exsol2} \ea
Certainly \re{exsol1}) is valid only  when it is meaningful; the
condition in \re{exsol2}) can be deduced directly from \rp extrem2 and
the fact the  considered extremal is nonnull.
It is straightforward to see that for the extremals \rp exsol1
\be ds={1\over\sqrt{|c-h|}}dx^0 \, ,  \el 11
while the 
extremals \re{exsol2}) are obviously always complete.

\section{Penrose Diagrams from Gluing}

\label{secPen}

 In this section we will provide the general rules of how
to find the Penrose diagrams starting from any given metric of the form
\re{metric}). This shall be done by  means of a generic example, the
function $h$ of which is drawn in Fig.\ 2a.
Having derived a simple building block principle, we will apply it to
the Jackiw-Teitelboim model (deSitter gravity) (\ref{JT}), $R^2$-gravity
(\ref{R2}), and the Katanaev-Volovich model (\ref{KV}) at the end of this
section.

In Fig.\ 2b  we qualitatively depicted  representatives of the null extremals
\re{null2a}, \ref{null2b}) corresponding to the metric of Fig.\ 2a.
 Any other solution
\re{null2a}) is obtained by shifting these curves along the
$x^1$-direction, a consequence of the fact that ($x^0=$ const)--lines are
Killing (isometry) directions within the chart $x^\m$.
The other type of null lines,
\re{null1}), are parallels to the $x^0$-axis; from them we have drawn only
those two which are the asymptotics to the representatives plotted
of the former kind in the sector on the right.

It is well-known that any  metric in two space-time dimensions is conformally
flat, i.e.\ that by a change of coordinates it can be brought into the form
$g=\exp (f(x)) dx^+dx^-$; this can be achieved even on a global level,
if the space-time manifold may be covered by one chart (and has  no closed
timelike curves). Within any of the sectors of Fig.\ 2b the diffeomorphism
\be x^+=x^1 + f(x^0),\, x^- =x^1 \el conf
with
\be f(x^0) \equiv 2 \int^{x^0} {du \0 h(u)}  \el fun
provides such a transformation; it, however, breaks down at $h(x^0)=0$.
It may be difficult to write down explicitly the diffeomorphism that
brings $g$ into conformal form on all of the chart underlying Fig.\ 2b.

Fortunately, the explicit form of such a diffeomorphism need not be
constructed. Similarly as in \re{conf})
it will be possible to choose $x^1$ as one of the light cone coordinates.
The diffeomorphism will then have the effect of straightening the null
extremals \re{null2a}), leaving \re{null2b})  as well as \re{null1},
i.e.\ $x^1=x^- =
\hbox{const}$, unmodified. Note that the $(x^+, x^-)$-chart cannot be all of
$\dR^2$ anymore; rather on the righthand side there will be some
boundary because the null lines of type \rp null2a do not intersect all
null lines $x^- =$ const. By means of a subsequent conformal diffeomorphism
$x^- \to \tan x^-$ (and a similar one for $x^+$)
the new coordinate chart covers only a finite region in $\dR^2$; the
result is drawn qualitatively in Fig.\ 2c. As indicated by the arrows,
the boundary on the righthand side can be made straight by a conformal
transformation in $x^+$, by means of which one can also transform all
rectangles into squares. The final building block for the Penrose diagram
is obtained by turning the patch 45 degrees counter-clockwise (given
our convention that a positive (negative) $ds^2$ corresponds to a
timelike (spacelike) distance) and is depicted in Fig.\ 2d.

In the Figs.\ 2c,d we also included the
lines of constant $X^3$ (= Killing directions), which in Fig.\ 2b
have been the straight lines $x^0=$ const.   The function
$X^3$ increases monotonously from the lefthand side,\footnote{For simplicity
we assume $\a =0 \Rightarrow x^0 =X^3$ here.}  where it is $-\infty$,
to the singularity on the righthand or upper side, where it gets $+\infty$.
In our example Fig.\ 2a both of these lines $X^3 \to \pm \infty$ are
complete, as the coordinate $x^0$ is unbounded in Fig.\ 2a (cf.\ the
discussion following Eq.\ \re{null1}));  completeness will be indicated
by boldfaced lines, incomplete singularities by thin solid lines (somewhat
thicker than the lines of isometry), and
horizons by dashed lines.

The square-shaped or triangular sectors are the regions with $(X)^2\neq 0$
($\LRA h \neq 0$). As already noted above the
triangular shape of the rightmost sector is due to the fact that 
any  type \rp null2a null extremal approaches a type \rp null1 extremal 
assymtotically for
$x^0\to+\infty$ (which implies that the former kind of null lines does 
not intersect all of the latter kind). 
If we had instead $\int^{+\infty} {dx^0\over h}=\pm\infty$
(e.g., $h=O(x^0)$),
all null extremals would intersect each other exactly once and the
triangular sector would be replaced by a quadratic one. If, on the other
hand,
e.g.\ $h\sim{(x^0)}^{n>1}\hbox{\ for $x^0 \to -\infty$}$, then,
vice versa, we would have to replace the quadratic sector on the
left of Figs.\ 2c,d by a triangular one, again with a timelike
boundary $X^3=-\infty$. And so forth.

Finally, in Figs.\ 2b,c
we have drawn some of the nonnull extremals \rp exsol1 with c=0,
i.e., \be {dx^1\over dx^0}=-{1\over h}, \el 13 as dotted lines. They
are the unique extremals
running through the cornerpoints of the $(X)^2\ne 0$ sectors in all nonnull
directions. Their length follows from \re{11}):
$$ s=\int_{\x_1}^{\x_2} {dx^0\over\sqrt{|h|}} $$
where $\x_1$, $\x_2$ are the successive zeros of $h(x^0)$.
Since in the chart \rp metric these extremals differ only by a shift in
$x^1$-direction, they all have the same length. It is finite only at single
zeros of $h(x^0)$:
\be s=\int {dx^0\over\sqrt{|h|}} \sim \int^\xi
{dx^0\over{(x^0-\xi)^{n\over2}}} \to \left\{
                         \begin{array}{r@{\quad}l}
                                <\infty & n=1 \\
                                 \infty & n\ge 2
                         \end{array} \right. \quad. \el length
Thus the cornerpoints of a sector between two single zeros of $h(x^0)$
are conjugate points.

Now any region with $(X)^2\neq 0$ can be found in two charts of the kind
(\ref{metric}),
one with $X^+\neq 0$ and one with $X^-\neq 0$, and it must be possible to
establish a diffeomorphism (at least locally) between them.
Such a diffeo\-morphism has to maintain $X^3$ and must map type \rp null1
null extremals onto type \rp null2a null extremals and vice versa.
By this it is already uniquely determined as
\be {\widetilde x^0}  =x^0 \quad , \qquad
\widetilde x^1= -f(x^0)-x^1+\hbox{const}
\el 12a
where $f(x^0)$ is the function \rp fun introduced already above,
and the integration constant
has been written down explicitly, indicating the free choice of the origin
of the $x^1$-coordinate.
It is easily seen that this diffeomorphism maps the whole sector where
$(X)^2\neq 0$.
A subsequent Lorentz transformation is necessary to restore our gauge:
\be e^+   \rightarrow -{2\over h}e^+ \qquad      e^-   \rightarrow
-{h\over 2}e^- \qquad    \omega \rightarrow \omega-d(\ln \! \!
\mid \! h \! \mid).
\el 12b
The final result of this transformation (\ref{12a}, \ref{12b})
is exactly that of \re{5}).
In the Penrose-diagrams the sectors are the squares (or the triangles),
the second solution is obtained from the first by taking the mirror image
(around an axis running diagonally through the sector, transversal to the
Killing-direction!) and the gluing diffeomorphism \re{12a}, \ref{12b})
amounts to
patching the corresponding sectors together. Fig.\ 3 illustrates
this process. By the above description gluing is unique, up to
the constant, which, as long as only the universal covering is
pursued, does not affect the solution.

We have yet to investigate the case (see Fig.\ 3)
that after surrounding the  point at the vertex of four diagrams
the corresponding sectors of the first and the fourth diagram
match. Shall they be identified?

The null extremals running into this vertex point are of the
type \re{null2b}).  Now, if this zero of $h(x^0)$ is simple
($h'(x^0)\neq 0$) then the answer is yes: The transformation
$$ \widetilde x^0=x^0 \quad \widetilde
x^1=x^1+{f(x^0)\over2} \, , $$
 with $f(x^0)$ as before,  which
is on each $(X)^2\neq 0$ sector a diffeomorphism, brings the
metric into Schwarz-schild form $$ ds^2=-{1\over h(\widetilde
x^0)}(d\widetilde x^0)^2 +h(\widetilde x^0)(d\widetilde x^1)^2,
$$ and the Kruskal extension (cf.\ e.g.\ \cite{Haw}),
plus simultaneous Lorentz transformation,
reveals the vertex point as regular interior point (a
saddle point of $X^3$, with $X^a=0$). The four adjacent
sectors then constitute one single sheet.

However, for zeros in $h(x^0)$ of a higher degree this procedure fails.  And,
in fact, not only the null extremals \ref{null2b} but also the general
extremals \ref{13} running towards this point are then complete
(cf.\ Eq.\ \ref{length}).
Thus the vertex point has to be taken out
of consideration. Also, to obtain the universal covering we must not identify
the diagrams but continue the gluing indefinitely.


The principle of how to construct {\em the} Penrose diagram corresponding
to any function $h$ in \rp metric should be clear now. The number (and
kind) of zeros of this function determines the number of squares in a
fundamental building block,
\footnote{The case that $h$ has no zeros at all
will be discussed in the applications below.}
the end of which is either a square or a
triangle, depending on the asymptotic behavior of $h$. The complete
Penrose diagram is then obtained by straightforward (pictorial) gluing.

Let us come to the announced examples, starting with deSitter gravity
(\ref{JT}). Since all values $\L \neq 0$ yield (basically) equivalent
Penrose diagrams, we will set $\L:=1$ in the following.
We then get a one-parameter family of functions $h$, $h(x^0)=C - (x^0)^2$,
parametrized by the Casimir constant $C = (X)^2 + (X^3)^2$ (cf.\ Fig.\ 4).

For any $C>0$ this curve $h(x^0)$  has two simple zeros, leading to
one square within the fundamental building block.
Asymptotically we have $h\sim-{(x^0)}^2$, so that
adjacent to the square there will be a triangle at each side, the boundaries
 of which, $X^3 =  \infty$ resp.\ $X^3 = - \infty$,  are spacelike and
complete. Gluing leads to the ribbon-like diagram shown in Fig.\ 5a.
(For the other sign of the cosmological constant, $\L=-1$, we get the same
diagram for $C<0$, rotated, however,  by 90 degrees, as the infinity is
timelike then).

For $C=0$ we get no square, but only two triangles. The
corresponding Penrose diagram is plotted in Fig.\ 5b.

For $C<0$ the function $h$ has no zeros (Fig.\ 4). We therefore may
apply directly the diffeomorphism (\ref{conf})  with the function
(\ref{fun}), which in the present case can be written  in terms of
elementary functions:
\be f(x^0)=-{2 \0 \sqrt{-C}} \arctan \left( {x^0 \0 \sqrt{-C}} \right) \, .
\el JTfun
The resulting region $(x^+,x^-)$ is again a ribbon (Fig.\ 5c);
but this time without any internal structure, as the Killing lines
$X^3=x^0=$ const become the parallels $x^+-x^-=$ const in the present
case.

It is straightforward to see that Fig.\ 5c depicts the Penrose diagram
for any (negative) function $h$ without zeros  which diverges at the infinity.
Clearly there is no conformal diffeomorphism
which maps this ribbon into a finite region. Let us remark also that
the Penrose diagrams for constant curvature in two dimensions are quite
different from the ones in  four dimensions (cf., e.g., \cite{Haw,Thi}).

Let us now turn to the second example: $R^2$-gravity. We have
already studied the behavior of the corresponding function $h$ when
discussing the quantum theory in Sec.\ \re{secgravaction}, cf.\ Fig.\ 1.
It is completely straigthforward to construct the  Penrose
diagrams according to the above rules. The result is depicted in
Fig.\ 6. For nonnegative $\L$ there are in addition to those diagrams also
infinte bands for the  constant curvature solutions $d\o = \pm 2 \sqrt{\L} \e$
(cf.\ Eq.\ \re{deSitter}); in the diagram Fig.\ 1 they are located at the
curve  {\bf R2,3,4}.

\medbreak The third example is the Katanaev-Volovich model \re{KV}). Its
potential is
\be V^{KV}=  (X^3)^2 -\L + {\a \over 2} (X)^2 \el VKV leading to the Casimir
function
\be C^{KV}= {2 \0 \a } \exp (\a X^3) \left( V^{KV} - {2 X^3 \0 \a} +
{2 \0 \a^2} \right] -{4 \0 \a^3} + {2 \L \0 \a} \, . \el CKV
From this and (\ref{h},\ref{gaugeB},\ref{h2}) one can determine the function $h(x^0)$.
The coordinate transformation $\a x^0 \to x^0$, $x^1 /\a \to x^1$,
constituting a residual gauge freedom of \re{metric}), combined with the
rescaling
\be \ti \L := \a^2 \L \, , \, \, \ti C := \a^3 C^{KV}  - 2 \ti \L + 4 \, \ee
brings it into the simpler form ($x^0 \in \dR_+$)
\be h(x^0) = {1 \0 \a}  \left\{ \ti C x^0 - 2 (x^0)^2 [(\ln x^0-1)^2+1-\ti
\Lambda] \right\} \, .\el hKV
It shows that, up to its sign, $\a$ does not influence the causal structure
of the KV-model.

The function $h$ always has a  zero at $x^0 =0 \LRA \a X^3 = -
\infty$, which is simple  for $\ti C \neq 0$. It corresponds to
an incomplete null-infinity.  On the other boundary of the
coordinate patch, $\a X^3 = + \infty$, we find $\a h \sim
(x^0)^2 \ln^2 x^0$, which  shows that the Penrose diagrams have
a complete triangular sector at this end.

To study the number and kind of zeros of the function \re{hKV})
for positive values of $x^0$, one is well advised to change variables
according to $y = \ln x^0$, being left with the {\em equivalent} analysis
of the number and type of zeros of the function $f(y) =
\ti C \exp (-y) -2[(y-1)^2 -\ti \L +1]$ within $y \in \dR$.
(Actually $y = \a X^3$ and $f(\a X^3) \equiv \a^3 (X)^2(X^3)$). In any case
the analysis yields
11 qualitatively different cases depending on the parameters $\ti C$
and $\ti \Lambda$:
\halign{{\bf G#}:\quad&#\hfill\cr
     1,2 &no zeros of $h$ and thus also of $(X)^2$\cr
      3  &one single zero at $X^3=B$                         \cr
      4  &one triple zero at $X^3=0$                      \cr
5,6 &one double zero at $X^3=\mbox{sgn}(\a) \, \sqrt\Lambda$                \cr
7  &one double zero at $X^3=-\mbox{sgn}(\a) \,\sqrt\Lambda$
and one single zero at $X^3=B_1$       \cr
8,9 &two single zeros at $X^3=B_2$ and $X^3=B_1$               \cr
10 &one single zero at $X^3=B_3$ and one double zero at $X^3=\mbox{sgn}(\a) \,
\sqrt\Lambda$ \cr
11 &three single zeros at $X^3=B_3$, $B_2$, and $B_1$,\cr}
where $B_{3(1)}<-\sqrt\Lambda<B_2<+\sqrt\Lambda<B_{1(3)}$ for $\a >(<)0$
and $-\infty<B<+\infty$. An overview is provided by Fig.\ 7.

Via \re{crit},\ref{V},\ref{VKV}) the critical values of $X^3$ are easily
determined to be $\pm \sqrt{\L}$; the corresponding value of the (rescaled)
Casimir function $\ti C$ is
\be \ti C_{crit} \equiv \ti C_{deSitter} = -4 \exp\(\pm \sqrt{\ti \L}\) \,
\( \pm \sqrt{\ti \L} -1 \) \, , \el CdeS
which marks the curve {\bf G5,6,10,4,7} of Fig.\ 7 and simultanously
the deSitter solutions $De^a=0, \, d\o = \pm  \sqrt{\L} \e$ (cf.\ Eqs.\
\re{momdeSitter},\ref{deSitter})). The cases {\bf G1,2}, as
well as {\bf G5,6} and {\bf G8,9}, differ by the kind of zero of $h$
at $x^0=0$; this has its impact on the completeness of the corner point.
It is now straightforward to draw the Penrose diagrams of the KV-model.
The result is depicted in Fig.\ 8 for $\a >0$; the diagrams for $\a <0$ are
obtained by rotating these by $90$ degrees.

The numbering {\bf G1-11} has been chosen as in \cite{Kat}, where the Penrose
diagrams Fig.\ 8 have been constructed first. It should be noted, however,
that our procedure to obtain these diagrams is incomparably faster than the
one of \cite{Kat}. The main  reason is that the local solutions
used in \cite{Kat} (resulting also from ours through the diffeomorphism
 \re{conf})) are valid only in coordinate patches which are part of ours
(the sectors $(X)^2\neq 0$); they had to be glued along their border, which
entailed lengthy considerations of the asymptotic behavior. In our gauge,
instead, the charts overlap and simply have to be matched together.
As a consequence we also could prove that all the solutions of \re{KV}),
(and in fact also of \re{gravaction},\ref{V}) with an, e.g., analytic potential $v$)
are analytic. Also, from \rp metric the existence of a Killing field is
immediate.

Concluding we remark that for many of the Penrose diagrams, such
as, e.g., for {\bf G3, G9}, it is possible to find also global
coordinates displaying explicitely the analyticity of $g$.
Examples for this might be given elsewhere.

\section{All Global Solutions and a Comparison with the
Quantum Theory:  The Example of the  Katanaev-Volovich-Model}


In the following we shall give an account of global solutions obtained by
factoring the universal coverings by a discrete transformation group.
In this way one obtains {\em all} global solutions. Among these are
the solutions with cylindrical topology. Keeping track of all diffeomorphism
(and Lorentz) invariant quantities characterizing such solutions, we get
some insight into the reduced phase space (RPS) of the theory as
defined on the cylinder. Note that the Penrose diagrams are labelled only by one
constant, namly the value of the Casimir function $C$ (beside, of course, the
coupling constants fixed in the Lagrangian). Since the result of a symplectic
reduction is  again  a symplectic space, with some care we will be able
to find a second continuous parameter resulting from the compactification.

In the  sections \ref{secquan} and \ref{secgravaction} we have studied
already the quantum theory of the space of gauge-inequivalent solutions
with cylindrical topology.
Certainly, a comparison of this quantum theory with the classical space-time
manifolds that are subject to this quantization is  worth an
investigation. The comparison  will be seen to  give rise to arguments in favor
of the quantization scheme employed, but there will arise also
 arguments questioning it. On the one hand we will find that
the topology of the RPS fits quite perfectly to the arguments of the wave
functions \re{wavfunc}); in particular, the quantum numbers $n_C$ are
in one-to-one correspondence with the minimal number of building blocks
intersecting a noncontractible loop on the cylinder. On the other hand, a
considerable portion of the solutions in the RPS  are incomplete, they have
closed timelike curves, and for some of them
it would be more natural to be reckoned
among other space-time topologies than the cylindrical one.

Most of these questions will be analyzed at the example of the
Katanaev-Volovich model (\ref{KV}),
but the discussion transfers in an obvious
way to the general model \re{gravaction},\ref{V}).

Let us start with classifying all possible 'compactifications' of
the Penrose diagrams. They are obtained by factoring out a discrete
transformation group from the universal covering solutions. Any such transformation
must of course preserve the functions $X^3$ and $X^a$.  Hence the sectors with
$(X)^2\neq 0$ must be mapped as a whole onto corresponding ones (i.e., with the
same range of $X^3$).

Within such a sector we have already discovered a Killing field
($\partial\over\partial x^1$ in the coordinates (\ref{metric})).  The
transformation generated by it is in local charts (\ref{metric}) a shift of a
certain amount in the $x^1$-direction. The gluing diffeomorphism (\ref{12a},
\ref{12b}) shows that such a transformation extends uniquely onto the whole
universal covering, and that it is in all charts represented as an $x^1$-shift
of the same amount (but on part of them in the opposite direction!). We will
call these transformations simply 'Killing-shifts'. In the Penrose diagram such
a Killing-shift shows as a distortion along the lines of constant $X^3$.

A further transformation of a sector onto itself is exchanging the two types
of null-extremals. This is exactly the gluing diffeomorphism of (\ref{12a},
\ref{12b}). It can also be described as a 'reflection' at one of the
extremals \re{13}).

Any admissible transformation can thus be separated into a combinatorial
part --- a certain permutation of the $(X)^2\neq 0$ sectors and their possible
reflections --- and one real parameter describing the Killing-shift. It is also
true that any transformation of the universal covering is already fully
determined by the image of only one sector.  The remaining investigations
shall be performed  ad hoc at the example of the Katanaev-Volovich
model now.\footnote{The
numbering below refers to the numbering of the Penrose diagrams in the
Figs.\ 7,8.}

\medskip
\noindent{\bf G1,2:} The only transformations are reflections and Killing-shifts.
Since a reflection has fixed points (an extremal of type \re{13}) as
symmetry-axis) it has to be ruled out.  The only discrete subgroups of
Killing-shifts are the infinite cyclic groups generated by one shift. The factor
space is then clearly a cylinder. In the coordinates (\ref{metric}) it can be
obtained by cutting out a strip parallel to the $x^0$-axis and gluing it
together along the frontiers. The width of this strip (i.e., of the generating
shift) is proportional to the length of a constant curvature path running once
around the cylinder. (Remember that in this model $X^3 \propto$ Ricci scalar,
on-shell). Hence for any value $C \ge 0$ we get a set of distinct solutions
parametrized by their size (any positive real number).

\medskip
\noindent{\bf G3:} As before reflections and also the inversion at the central
saddle point must be dropped. We could again try a cylinder-solution, but there
occur some problems, if one adopts a standard gravity point of view:
Not only is there a closed null-extremal
($x^0=\hbox{const, at $(X)^2(x^0)=0$}$), but other extremals approach it
asymptotically, winding around the cylinder infinitely often while having only
finite length. This situation resembles strongly the Taub-NUT space or rather
its two-dimensional analog as described by Misner
\cite{Mis} (cf.\ also \cite{Haw}). As explained there an extension
is possible, if one abandons the Hausdorff property. The net result can be
described as two concentric cylinders attached to each other at the (closed)
line $(X)^2 =0$; those extremals which previously had been incomplete are now
continued at the 'other sheet' previously not included.  In fact, the
resulting extended solutions occur naturally as factor space of the universal
covering.

On the other hand, from the purely field theoretic point of view, there is no
notion of geodesic (or extremal) completeness; the
solutions on the cylinder are perfectly analytic everywhere on the
cylinder (cf.\ Eqs.\ \re{metric},\ref{gaugeA},etc.)
taking the $x^1$-coordinate as periodic now),
and there also is the metric induced circumference as the variable
'conjugate'  to $C<0$. (Changing the length of periodicity
of $x^1$, one can change the length of an ($X^3=$ const)--line; since such a
circumference is, for any fixed value of $X^3$, a gauge-independent quantity,
it may represent the second variable besides $C$).

Taking together the cylindrical solutions for {\bf G1-G3}, from the field
theoretic point of view we find a perfect
coincidence between the quantum theory of the KV-model with $\L <0$
(cf.\ Fig.\ 7) and the corresponding RPS, which is just a plane
(identifying the four possible cylinders for $C<0$ by means of the
discrete symmetry transformation \re{5})). Nevertheless, according to
the above considerations, it seems more natural to regard the cylindrical
solutions for $C<0$ as a {\em part} of the extended nonHausdorff  object
described above. A possible, somewhat speculative
interpretation of the quantum theory for $\L <0$
could then be that wave functions with support on $C \ge 0$ correspond to
cylindrical space-times, their counterparts with $C<0$
correspond  to these 'four-sheet' space-times, and quantum processes mixing
these vector-subspaces  describe topology changes. The other way out might be
to just regard the  Hamiltonian methods used so far to be  not sophisticated
enough to cope with such problems of quantum gravity as geodesic completeness
and different topologies.

In the following we will continue studying all  factorizations possible in the
remaining solutions {\bf G4 - G11}.
Since Taub-NUT like solutions
exist for {\em all} of these cases,  we will not mention them any further.

\medskip
\noindent{\bf G8,9:} Since this solution is an
infinite ribbon of equal building blocks, it is possible to factor
out a shift of a number of blocks to obtain a cylinder.
Furthermore, while pure reflections (and inversion at a saddle
point) have to be ruled out, a 'vertical' reflection plus 'horizontal' shift
will work and it yields a M\"obius-strip.

Let us now come to find the
second phase space parameter (for the cylindrical solutions).
It was already pointed out that the generating shift-transformation has an
additional real parameter, the Killing-shift component. We have yet to show
that different Killing-shift values yield inequivalent solutions:
In the previous section it was proved that the saddle points are conjugate
points and the extremals running between them are those of \re{13}). They run
through the saddle points in all directions between the two null-directions.
A Killing-shift shifts them sidewards, altering the angle of their tangent.
One can now start from a saddle point in a certain direction along a spacelike
extremal. This extremal will eventually return to the original point, but due
to a Killing-shift its tangent at the return  may be tilted against that at the
start. Since this tilt can
be expressed in terms of a shift of the $x^1$-coordinate in the
chart \re{metric}),  it is the same for {\it all} such
extremals. Especially, there is one solution without tilt.
Thus, besides the parameter $C$, the cylindrical solutions are
parametrized by a positive integer (number of copies or
`building blocks') and a real constant parametrizing the
tilt.\footnote{If we had chosen $\a <0$ the whole Penrose diagram
would have to be rotated by 90 degrees.  The above extremals
would then be timelike and the tilt at the return could be
interpreted nicely as acceleration along one journey around the
cylinder.}

It may seem that for the M\"obius-strip one might also have this continuous
parameter.  However, in contrast to the former examples sectors are
occasionally identified with their mirror images. 
This has the consequence that on the factor space a Killing-shift cannot
be defined consistently. Furthermore there is exactly one extremal which has
the same tangent even at the first return. A Killing-shift component of the
transformation only results in a different choice of this special extremal and
thus (by a coordinate change) leads to equivalent solutions.  Hence, besides
$C$, the M\"obius-strip solution is only parametrized by a positive integer
(number of copies).


\medskip
\noindent{\bf G5,6:} Again reflections cannot be used. As for {\bf G8,9}
we get
cylinders parametrized by a positive integer (the number of copies) and a
real number (Killing-shift),
but this time no M\"obius-strips.
In contrast to G8,9 there is no such nice description of the Killing-shift
parameter, because we have in general no closed extremals. One can, however,
take a series of null extremals, zigzagging around the cylinder between two
values of $X^3$ (or in this case even an oscillating spacelike extremal) and
interpret the failure to be closed (i.e., the distance between starting- and
endpoint on this $X^3=\hbox{const}$ line) as measure for the Killing-shift.

\medskip
\noindent{\bf G4:} As before this solution is an
infinite ribbon of equal building blocks (although this time not straight but
winding around the central point). Thus one obtains again cylinders of a
certain number of copies.
When passing once around such a cylinder, however, the light cone tilts upside
down $n$ times, where $n$ is the number of copies involved.  Thus we have got
an {\it$n$-kink\/}-solution. Hence the nonTaub-NUT solutions are cylinders
parametrized by a positive integer (the number of
copies = number of kinks) and a real number (Killing-shift).

\medskip
\noindent{\bf G7,10,11:} These cases are slightly more complicated. Evidently
G7 and G10 give rise among others to a cylinder with hole(s) and G11 to a
torus with hole(s). Furthermore, each hole yields an additional real
parameter (characterizing the re-identification after surrounding this hole)
such that the $n$-hole cylinder or the $n$-hole torus has $n+1$ real
parameters.

But even a series of proper (yet slightly pathological)
cylinders can be obtained:
The universal covering covers the cylinder resp.\ the torus with hole. The
group of cover-transformations is isomorphic to the fundamental
group of the covered space, in this case the free group with two generators.
It contains only admissible (isometric etc.) transformations but not all of
them (e.g.\ reflections are missing). Any subgroup yields a factor space,
especially a cyclic subgroup yields a cylinder.

To speak in pictures: The generator of this cyclic subgroup defines a path
in the universal covering. Now the end-sectors of the path (i.e.\ of the
corresponding ribbon) are identified and at all other junctions the solution
is extended infinitely without further identifications. Thus a topological
cylinder (although with a terribly frazzled frontier) is obtained.

For reasons of completeness one should
treat also the deSitter solutions \re{momdeSitter},\ref{deSitter})) of the KV-model,
corresponding to the Casimir values \re{CdeS}). Partially this gap will be
closed in the next section, where we will  concentrate  on
all cylinder-solutions of the Jackiw-Teitelboim model. Since, however, the
momenta are constant all over the space-time manifold in this case, there
might be additional possibilities to compactify the deSitter solutions of the
KV-model as compared to the ones  found in the regular sector of the JT-model.
We shall not investigate this here further.

Note also that from the quantum theory point of view the deSitter solutions,
as well as the cylinder solutions {\bf G4,5,6,7}, and {\bf G10},
 should be more or less negligible as  they  correspond to only one
value of $C$ (cf.\ also Fig.\ 7). The  space of orbits, furthermore,
is nonHausdorff exactly at these solutions.

Again the discrete indices present in \rp wavfunc fit to the cylinder-solutions
found above. However, at least in the case of {\bf G11}
I feel some unease with these 'cylinders'. And now one also cannot reinterpret
the wave functions of this sector in the RPS
to actually describe the natural factor space of a torus with hole, since the
latter is parametrized by {\em three} continuous quantities.

\section{Symmetries,  Metric-Nondegeneracy, and Kinks: The Example of the
JT-Model}

\label{secJT}

One of the features of the Ashtekar formulation of 4D gravity
usually considered as an
advantage is that the formulation is well-defined also for configurations
corresponding to degenerate metrics. In particular, the field equations and the
symmetry transformations are equivalent to the usual Einstein formulation only
for $\det g \neq 0$. It is the purpose of this section to investigate a similar
relationship between two formulations of the symmetry content of the
class of gravity models considered in this work which are also
equivalent for nondegenerate configurations only.

According to \re{diff}) and \rp identi the  gravitational symmetries, i.e.\
diffeomorphisms and local Lorentz transformations, can be identified with the
symmetry transformations \re{sym2a}, \ref{sym2b})
 on-shell, if and only if $e :=$ det$e_\m^a
\neq 0$ ($\LRA \det \, g \neq 0$).  As  is obvious from  the paragraph of Eq.\
\re{consalg}) and the one following it, the symmetries \re{sym2a},
\ref{sym2b})
coincide with
the Hamiltonian symmetries; factoring out the former from the space of
solutions to the field equations (for cylindrical space-time topology) is
equivalent to a symplectic reduction\footnote{A symplectic reduction is
performed by implementing a system of first class constraints $G^i=0$ in a
phase space,  factoring out the Hamiltonian flow of the $G^i$ on
this subspace subsequently.}   yielding the reduced phase space (RPS)
of the theory.
The latter was the space subject to  quantization in sections
\ref{secquan} and \ref{secgravaction}.

In this section let us  compare the following two moduli spaces:

\smallskip
{\bf 1)} The space of  ($C^\infty$-)solutions to the field equations on
the cylinder modulo the symmetry transformations \re{sym2a}, \ref{sym2b}).
As noted above this space is equivalent to the standard RPS.

\smallskip
{\bf 2)} The same space of solutions, but excluding from it all configurations which
correspond to a somewhere degenerate zweibein; only those solutions are
identified which are related by gravitational symmetries.

There are basically two  reasons that could give rise to a difference between
these two spaces. Firstly, there could exist gravitationally inacceptable
solutions to the field equations which are not gauge related to any
gravitationally acceptable one. In this case the RPS of {bf 1} contains points not
included in {\bf 2}. Secondly, symmetry orbits of {\bf 1} could be cut into pieces by
regions (in the space of solutions) which have somewhere (in $M$) degenerate
zweibeins.  In this case there are several points in {\bf 2} corresponding just to
one point in {\bf 1}.

The first of these two mechanisms occurs for  the Euclidean formulation of the
gravity theories. Only bundles which have a Chern class that coincides with the
one of the canonical (tangential) bundle on $M$ will yield nowhere degenerate
metrics. E.g.\ on a sphere the trivial bundle will in no way yield
nondegenerate metrics; one necessarily will have to introduce at least two
charts with a nontrivial gauge transformation
\re{sym2a}, \ref{sym2b}) on their overlap.

The second mechanism occurs in the case of Minkowskian gravity  on $M=S^1
\times \dR$ and  shall be illustrated by means of a simple example: Take
on the one hand the real line $\dR$ ($\sim$ space of all solutions) and as the
symmetry transformations  translations with generator $T_1=\partial /\partial
q$ ($\sim$ symmetries \re{sym2a}, \ref{sym2b})).
 Take on the other hand $\dR - \{0\}$ ($q=0
\sim e=0$) modulo the transformations generated by $T_2=qT_1$ ($\sim$
gravitational symmetries). The degenerate point $q=0$ is gauge related to $q
\neq 0$ with respect to $T_1$; thus we do not have a problem of the first kind.
However, the symmetry orbit of $T_1$, which reduces $\dR$ to a point, is cut
into two pieces by the fixed point $q=0$ of the $T_2$-transformations.  On {\em
all} of  $\dR$ there are three gauge orbits of $T_2$,  $\dR^+$, $\dR^-$ and
$\{0\}$, corresponding to three point in the factor space.  Even if the point
$q=0$, where the correspondence between the infinitesimal form of the two
symmetry transformations breaks down, is  now removed, we end up with different
results.

We will show in the present section that indeed eliminating the solutions with
$\det g=0$, the gauge orbits of \re{sym2a}, \ref{sym2b}) split into components not smoothly
connected to each other. Solutions from different components of the same gauge
orbit are not related by gravitational symmetry transformations (since
obviously $\det g=0$ is a fixed point under the latter). They correspond to
space-time manifolds with different kink-number. Although it is possible to
parametrize the gauge orbits of the constraints globally for all models
\re{gravaction}, \ref{V}), we will restrict the analysis below to the case
$V=X^3$, i.e.\ to the Jackiw-Teitelboim model, in which case this
parametrization is particularly simple. As already noted in the introduction
for   this $V$ the action \re{gravaction}) can be rewritten
{\em identically} as \re{gaugeth2}) with Lie algebra
$so(2,1) \sim sl_2$ \cite{Isl}; thus
the gauge orbits are the standard  ones of a nonabelian gauge theory.

The Penrose diagrams of the Jackiw-Teitelboim model, Fig.\ 5, do not allow for
any complete kink solution. This coincides with the fact that all the kink
solutions we will obtain as representatives of the different  parts of the
gauge orbits cannot be geodesically completed. Nevertheless, from the field
theoretic point of view they constitute perfect $C^\infty$ solutions on $M$
(with an everywhere nondegenerate metric), which cannot be transformed into
each other by means of the gravitational symmetries but  are gauge related by
$sl_2$ transformations. The fact that these incomplete kink solutions are
eliminated automatically when using the Hamiltonian constraints $G^i$ as
symmetry generators
could be regarded as an advantage of the formulation
over  some other formulation which strictly implements only gravitational
symmetries. On the  other hand this coincidence could be regarded also as
purely accidental, two mistakes cancelling each other, an argumentation that
can gain some support from the observations made in the preceding section.

The investigation below provides also some insight into the topology of the
RPS. In particular it will be seen to be not Hausdorff. We suppose that this
inevitably leads to some ambiguity in the quantization (in addition to the
choice of polarization).

In the following we will determine the
Hamiltonian RPS by means of the equivalent group theoretic formulation, which
allows us to use comparatively simple fiber bundle methods. Since the
constraints can generate only gauge transformations connected to the identity
and since large gauge transformations are in one-to-one correspondence  to the
first fundamental group $\Pi_1(G)$ of the gauge group $G$, the gauge group $G$
we have to use is the universal covering group of $SL(2,\dR)$, denoted by
$\widetilde{SL}(2,\dR)$. The latter, however,  has no faithful (finite
dimensional) matrix representation. This technical obstacle is overcome by
first determining the factor space using the gauge group  $PSL(2,\dR)
\sim SL(2,\dR)/\{ 1,-1 \}$,  withdrawing then the additional identification
by applying large gauge transformations to the obtained representatives.  In
this way we will obtain representatives of the $\widetilde{SL}(2,\dR)$ gauge
theory, which from the group theoretical point of view describe equivalence
classes of paths in $PSL(2, \dR)$ parametrized by $x^1$.
All of these will correspond to solutions with $e \equiv 0$.
Parametrizing the $\widetilde{SL}(2,\dR)$-orbits through these
representatives, we then will find an infinity of gravitationally
inequivalent ways to ensure $e \neq 0$ everywhere on the cylinder $M$.

For reasons of explicitness let us choose a basis $T^i$, $i \in \{+,-,3\}$,
of the $sl_2$-algebra which satisfies $[T^-, T^+] = T^3$, $[T^\pm,T^3]=\pm
T^\pm$. A real matrix representation of this is provided by $T^\pm =
(\s^1 \pm i \s^2)/2\sqrt{2}$, $T^3=-\s^3/2$, where the $\s$'s are the standard
Pauli matrices. In this way  we can represent all fields by matrices
through $A=A_iT^i$ and $X=X_i T^i=X^iT_i$, where the indices shall be raised
and lowered  by means of half of the Killing metrik $\k$: $\2 \k_{+-}
= \eta_{+-} = 1, \2 \k_{33}=1$. With \rp identi one thus has
\be
A=\left( \begin{array}{cc}  -\o/2 & e^-/\sqrt{2} \\ e^+/\sqrt{2}& \o/2
\end{array}\right), \quad X=\left( \begin{array}{cc} - X^3/2 & X_+/\sqrt{2}
\\ X_-/\sqrt{2}&X^3/2 \end{array}\right) \, ,  \el matriidenti
and the action \rp gaugeth coincides with \rp gravaction for $V+X^3$.
The factors two and $\sqrt{2}$ above have been introduced so as to avoid any
conflict with the conventions chosen in the gravity formulation.
Note also the
following identity for the Casimir invariant: $C= X^iX_i =2tr(XX)=-4detX$.

The group $\cal G$ of the symmetry transformations we consider in the
first stage is the group of smooth mappings
from the cylinder into $PSL(2,{\bf R})$:\footnote{There are no nontrivial
principal $G$-bundles on a cylindrical base manifold, iff the  chosen structure
(gauge) group $G$ is connected.}
\be
{\cal G}_{PSL(2,\dR)}=\{g:S^1\times {\bf R}\to PSL(2,{\bf R})\}
\el gaut
The equations of motion, which in the matrix notation introduced above take the
form
\be
F=0, \qquad d X+[A,X]=0 , \la{Feom} \ee yield the connection to be flat and the
field $X$ to be covariantly constant. Up to gauge transformations a flat
connection $A$ on a cylinder is determined by its monodromy $M_A={\cal
P}\exp\oint A
\, \in PSL(2,{\bf R})$ generating parallel transport around the
cylinder ($\cal P$ denotes path ordering and the integration runs over a closed
curve $\C$ winding around the cylinder once).  As the exponential map is
surjective on $PSL(2,\dR)$, any monodromy matrix can be generated by a connection
of the form $A=A_1dx^1$ where $A_1$ is constant:
\be
A=\left( \begin{array}{cc}  z &y+t \\ y-t&- z
\end{array}\right)dx^1, \quad t,y,z \in {\bf R}   \, .\la{m} \ee
Constant gauge transformations act on $A$ via the adjoint action leaving the
determinant $t^2-y^2-z^2$ invariant and may be interpreted as Lorentz
transformations in the three dimensional Minkowski space $(t,y,z)$.
Hyperbolic, elliptic and parabolic elements, respectively, in the Lie algebra
correspond to spacelike, timelike, and lightlike vectors, respectively, in this
Minkowski space.  By Lorentz transformations in the $(t,y,z)$ plane they can be
brought into the form:
\be
\begin{array}{c}
A^{hyp}=\left( \begin{array}{cc}  0 & \a \\
\a& 0 \end{array}\right)dx^1, \quad
A^{ell}=\left( \begin{array}{cc}  0 & \dh\\ -\dh& 0
\end{array}\right)dx^1, \\ A^{par}=\left( \begin{array}{cc}  0 &
0 \\
\pm 1 & 0 \end{array}\right)dx^1
\end{array}\la{Arep} \ee
with $\a,\dh \in {\bf R} $ and the identification $\a\sim -\a$.  Exponentiation
yields the monodromy matrices
\be \begin{array}{c}
M_{A^{hyp}}=\left( \begin{array}{cc}  \cosh 2\pi\a & \sinh 2\pi\a \\
\sinh 2\pi\a& \cosh 2\pi\a \end{array}\right), \quad
M_{A^{ell}}=\left( \begin{array}{cc} \cos 2\pi\dh  & \sin 2\pi\dh\\ -\sin
2\pi\dh& \cos 2\pi\dh \end{array}\right), \\ M_{A^{par}}=\left(
\begin{array}{cc}  1 &  0 \\
\pm 2\pi & 1 \end{array}\right) \, ,
\end{array}\la{g} \ee
inducing the further identification $\dh\sim\dh +1/2$ in the elliptic sector
(remember $\oint dx^1=2\pi$ and $1 \sim -1$).  The integration of  the second
Eq.\ \re{Feom}) gives $X(x^0,x^1)=X(x^0,x^1+2\pi)=M_A X(x^0,x^1){M_A}^{-1}$. Thus
choosing a connection from (\ref{Arep}), $X(x)$ has to commute with the
corresponding monodromy matrix and consequently with the connection itself.
Using \rp Feom again, one finds $X(x)$ to be constant. We obtain:
\be\begin{array}{c}
X^{hyp}=\left( \begin{array}{cc}  0 &c_1 \\ c_1& 0
\end{array}\right), \quad X^{ell}=\left( \begin{array}{cc}  0 &
c_2\\ -c_2& 0 \end{array}\right), \\ X^{par}=\left(
\begin{array}{cc}  0 &  0 \\ c_3 & 0 \end{array}\right)\, ,
\end{array}\la{mom}
\ee
where the $c_i$ are arbitrary real parameters. Note, however, that due to
$(\a,c_1) \sim (-\a,-c_1)$ the hyperbolic sector
of the $PSL(2,\dR)$-RPS is a cone.

In the case $A=0$ (corresponding to $\a =0$ or $\dh=0$, respectively, in
(\ref{Arep})) $X(x)$ is constant, too, but it is not restricted by its
commutator with the monodromy matrix.  It is, however, subject to constant
gauge transformations, as they leave $A=0$ invariant. Considerations similar to
those above show that also in this case gauge representatives of the solutions
are given by (\ref{mom}) with $c_3=\pm 1$ and the identification $c_1\sim
-c_1$.

(\ref{Arep}, \ref{mom}) with $(\a,c_1) \sim (-\a,-c_1)$ together with the
$A=0$-sector give a complete parametrization of the
reduced phase space of the $PSL(2,{\bf R})$-gauge theory.
As the configuration variable $M_A$ is compact in the elliptic sector
of the RPS, the corresponding conjugate variable $C$ will have a discrete
spectrum for $C < 0$ in the quantum domain;
furthermore, there will exist some $\Th$-angle
within this spectrum as the RPS is not simply connected. ($\Th$ will
label the irreducible representations of the fundamental group of the
RPS which is ${\bf Z}$; thus $\Th \sim \Th + 2\pi$). Indeed, these
expectations have been confirmed in \cite{9402}.

As already indicated above, the group of gauge transformations ${\cal
G}_{PSL(2,{\bf R})}$ is not connected; rather it consists of
an infinite number of components not smoothly connected to each other:
$\Pi_0({\cal G})=\Pi_1 (PSL(2,{\bf R})) = {\bf Z}$.  A complete set of
representatives for the components of ${\cal G}_{PSL(2,{\bf R})}$ is given by
\be g_{(n)} = \left( \begin{array}{cc}
\cos (nx^1/2) & \sin (nx^1/2)\\
-\sin (nx^1/2) & \cos (nx^1/2) \end{array}\right) ,\qquad n\in {\bf Z} .
\la{groupel} \ee
Parametrizing the phase space as in (\ref{Arep}) - (\ref{mom})
we also implemented these gauge transformations. The action of the group
elements $g_{(n)}$ on the connections (\ref{Arep}) gives in the hyperbolic
sector
\ba A^{hyp}_{(n)}&=& \left( \begin{array}{cc} \a\sin (nx^1)
 & \a\cos (nx^1) + n/2 \\
\a\cos (nx^1) - n/2 & -\a\sin (nx^1) \end{array}\right)dx^1
\nn
X^{hyp}_{(n)}&=&c_1\left( \begin{array}{cc} \sin (nx^1)  &  \cos (nx^1)  \\
\cos (nx^1) & -\sin (nx^1) \end{array}\right)
.\la{trsol} \ea An analogous result is obtained in the parabolic
sector. In the elliptic sector the $g_{(n)}$ generate a transformation $\dh \to
\dh + n/2 $.  They are responsible for the previous identification $\dh \sim
\dh + 1/2$, which now is removed again.

In this way we have found a complete parametrization of the RPS of the
Jackiw-Teitelboim model. It agrees perfectly with the quantum mechanical system
obtained for it: Obviously $X_{(n)}^{hyp}$ are representatives of the
first homotopy of the coadjoint orbit $C=4(c_1)^2 >0$ and
the integer $n$ can be chosen to coincide with the discrete index found in
(\ref{wavfunc}) obviously coincides with the integer $n$ present in the
parametrization of the hyperbolic sector ($C>0$) of the RPS.
Note also that we had to used (at least) two charts to depict the
Jackiw-Teitelboim wave functions
in the form \re{wavfunc}); they correspond to
the two signs of $c_2$ in \re{mom}) which are swallowed within $C=-4(c_2)^2$.

At $C=0$ the RPS
is not Hausdorff: the parabolic sector has no disjoint neighborhood with
the ($A=0$, $C=0$)--part of the RPS.  Thus there will be no unique way to
connect the qualitatively different sectors $C>0$ and $C<0$ of the quantum
theory.

As indicated previously the RPS  above agrees also with the cylindrical
factor spaces obtained from the Penrose diagrams Fig.\ 5: $n$ counts the number
of blocks before the identification and the monodromy matrices are phase space
analogues of the 'tilt' and the 'circumference' found as the second gauge
independent variable beside $C$ in the hyperbolic and elliptic sector,
respectively.

The simplest possibility to bring any of the representatives  above into a form
corresponding to a nondegenerate metric is provided by the gauge transformation
$\exp (x^0 T^+) = 1 + x^0 T^+$. (This is true except for
$A \equiv 0$ where two transformations are necessary).
As a byproduct we find global charts for the
Penrose diagrams Fig.\ 5 and its cylindrical factor spaces in this way.

To get some understanding of the $(e=0)$-structure of the orbits, let us
parametrize a general $\widetilde{SL}(2,\dR)$-element $g(x)$ as follows:
\be  g(x) =
\left(
\begin{array}{cc} e^\a & 0 \\ 0 & e^{-\a} \end{array}\right)
\left(
\begin{array}{cc} \cos \psi & \sin \psi \\ -\sin \psi & \cos
\psi \end{array}\right) \left( \begin{array}{cc} 1 & \c \\ 0 &
1 \end{array}\right)  \, \, ,
\label{grouppar}
\ee
where $\a$, $\psi$, and $\c$ are arbitrary, in $x^1$ {\em periodic}
functions of $x$. That this is a true parametrization can be seen by noting
that the group $SL(2,\dR)$ could be defined  as the group of basis
transformations in a two-dimensional vector space which leaves the area between
two basis vectors invariant; by the first transformation one can change the
angle between the two vectors, the second one rotates them, and the third one
allows to bring one of the two vectors to any given length. The transition from
${\cal G}_{SL(2,R)}$ to ${\cal G}_{\widetilde{SL}(2,R)}$ is performed when we
excluded quasiperiodic functions $\psi$.

Now one has to apply the general $\widetilde{SL}(2,\dR)$--gauge transformation
to any of the representatives of the RPS. Let us do this at the example of
the elliptic sector. We can set $\a \equiv 0$ for our purposes, since
the third transformation  corresponds to a Lorentz transformation in the
gravity frame bundle and hence it  leaves $e^- \wedge e^+= e d^2x$ unchanged.
We then obtain:
\ba
 (A^{ell})^g &=&(\dh dx^1 + d\psi) \left( \begin{array}{cc}
    \c & (1+{\c}^2) \\
-1 & -\c \end{array}\right) + \left( \begin{array}{cc}
    0 & d\c \\ 0 & 0 ) \end{array}\right) \nn
 (X^{ell})^g &=& c_2 \left( \begin{array}{cc}
    \c & (1+{\c}^2) \\
-1 & -\c \end{array}\right) \, .
\la{novang} \ea
As noted  above the choice $\psi \equiv 0$, $\c := x^0$ yields a
nondegenerate solution (for $\dh \neq 0$): With \rp matriidenti we find $\e
\equiv e^- \wedge e^+ = d\o= -2 \dh d^2x$.
Let us now analyze the more general transformation  provided by
\be
\c =
r(x^0)\cos(kx^1), \,  \quad
\psi = r(x^0) \sin(kx^1),  \quad k\in {\bf N}_0 , \la{para}
\ee
where $r$ is some function of $x^0$. We then find that $\e = d\o =
r' [\dh \cos (kx^1) + kr] d^2x$.  So the resulting metric and zweibein
will be nondegenerate on all of the cylinder, iff $r$ is a  strictly monotonic
function which, for $k \neq 0$, is bounded by $\vert \dh \vert /k$ from below.
A possible choice is, e.g.,  $r(x^0):= \exp(x^0)\,+2\vert\dh\vert$.

Despite the fact that the gauge transformations (\ref{grouppar}, \ref{para})
with $\a \equiv 0$ and $r$ chosen as above are smoothly connected to the unity
for arbitrary value of $k$, the solutions $(A^{ell})^g$ are gravitationally
inequivalent for different values of $k$. To prove this let us  choose a loop
$\C$ running around the cylinder once.  Under the restriction $\det g = -(\det
e)^2 \neq 0$ the components of the zweibein $(e_0{}^+, e_1{}^+)$  induce a map
$\C\sim S^1\to {\bf R}^2\backslash {\{0\}}$ characterized by a winding number
(not depending on the choice of $\C$).  Solutions with different winding
numbers cannot be transformed into each other by gravitational symmetries,
since they are separated by solutions with $\det e =0$.   For different values
of $k$ the solutions (\ref{novang}) have different winding numbers, which
proves our assertion.

This result generalizes to the other sectors of the theory: Solutions which are
gauge equivalent in the $\widetilde{SL}(2,{\bf R})$ gauge theory are not
equivalent in the gravity theory (as defined in item {\bf 2} at the beginning of this
section), if they have different winding number. Having found this
inequivalence between the factor spaces {\bf 1}
and {\bf 2}, we will not be interested
in investigating the latter any further. In particular we will not factor out
the large gauge transformations of the gravity theory; since the group of
diffeomorphisms and local Lorentz transformations consists only of a finite
number of components  not smoothly connected to each other,\footnote{They
differ by $x^0$- and $x^1$-reflection on the space time manifold and by parity
transformation and time reversal in the Lorentz bundle.} the inequivalence will
not be removed by them.

The winding number defined above is related to the kink number as defined in
\cite{kink} by means of 'turn arounds' of the light cone along noncontractible
loops. More precisely, winding number $k$ corresponds to kink number $2k$. (Odd
kink numbers \cite{kink} characterize solutions which are not time orientable.
Such solutions are not considered here).

All the kink solutions found above are  geodesically incomplete.  The prototype
is provided by $k=1$ since the solutions with $k>1$ are $k$-fold coverings of
it. For $k=1$ it is helpful to regard $r$ and $x^1$ as polar coordinates for
hypothetical cartesian coordinates $\c$ and $\psi$.  With the observation that
$X^3 =2c_2\c$ and an analysis of the null extremals it can be  seen that the
($k=1$)--solution \re{novang}) is the result of cutting out some  piece  from the
Penrose diagram Fig.\ 5c including some part of one boundary; after the cutting
procedure the open ends are identified again. More details shall be
provided in \cite{Klo}.

Such a cutting procedure  can be performed with any Penrose diagram. Taking the
corresponding $k$-fold covering will yield a 2$k$-kink solution. Thus such
solutions exist not only for the Jackiw-Teitelboim, but also for the other
models considered in this chapter. By construction their maximal extension is
geodesically
incomplete,\footnote{Except for diagrams such as $G4$ of Fig.\ 7.}
providing however gravitationally inequivalent
nondegenerate $C^\infty$ solutions with cylindrical topology.

It could be regarded as an advantage of the Hamiltonian formulation with
constraints \re{cons}) that the geodesically incomplete ($k \neq 0$)--solutions
are automatically identified with the ($k=0$)--solution. However, since also the
latter are not  complete for any choice of $C$ (cf.\ the discussion in the
previous section), this 'advantage' seems rather accidental. One of the lessons
to be drawn from the analysis of this section is: Equivalence of
symmetry transformations only up to $\det g =0$ in general
is of relevance for the factor
spaces, and thus for the quantum theory, even if  finally all
degenerate solutions are excluded.

\section{Considerations on the Issue of Time}

All the models considered within this work  can be reduced to quantum systems
of finitely many topological degrees of freedom.  Thus the question arises: Can
such models serve as toy models for a quantum theory of four  dimensional
gravity?  We hope to have convinced the reader  within the last sections  that
this is the case with respect to some technical questions arising in any theory
of quantum gravity. It is the purpose of the present section to show that also
an illustrative treatment of conceptual questions is possible. We will focus on
the  so-called 'problem of time' \cite{Ish} of quantum gravity, i.e.\ the
question of how to find any dynamics within such a  theory as the standard
Hamiltonian vanishes on all (physical) quantum states.

For this purpose we study the example of $R^2$-gravity with Minkowski signature
coupled to $SU(2)$ Yang Mills. The Lagrangian of this system is
\be L=  \int_{S^1 \times {\bf R}} [{1 \0 8\b^2} R_{ab} \wedge \ast R^{ab}
+{1 \0 4\g^2} tr(F \wedge \ast F)]  \la{S}    \ee where   the Hodge dual
operation is performed with the dynamical metric used to define also the
torsionless curvature two-form $R_{ab} = \e_{ab} \,d \o(e)$, and the trace is
taken in some representation of $su(2)$. We may, e.g., choose $T_i = \s_i/2i$
and use the metric $-2tr{T_iT_j}=\d_{ij}$ to lower and raise Lie algebra
indices.   Rewriting $\re{S})$ by means of Cartan
variables in a Hamiltonian first order form, it becomes \be L_H=\int_{S^1
\times {\bf R}} B_aDe^a+B_3d\o + tr(EF) - [\b^2 (B_3)^2-\g^2
tr(E^2)]\e \la{SH} \ee
where we have chosen $E =E^iT_i$ to denote the 'electric fields'
conjugate to the $SU(2)$-connection one-components $A_1$, and
the $B$'s are the conjugates to the spin connection $\omega_1$ and the
zweibein one-components $e_1{}^a \equiv  (e_1{}^-,e_1{}^+)$.

$B$ and $E$ together can be interpreted as coordinates $X$ in a six-dimensional
target space $N$ with an appropriate Poisson structure defined on this space.
In the present case it is, however, simpler to regard $B$ and $E$ as
coordinates for two three-dimensional Poisson structures.
$tr (E^2)=-E_iE_i/2$ may be seen to be  a Casimir function
of the six- as well as of the two three-dimensional Poisson structures. Thus
on-shell it is a constant. $S_H$ is the  sum of an
$SU(2)$-$EF$-theory \re{gaugeth2}) (up to a factor $-2$) and an action
 \re{gravaction}) with $V= \b^2 (B_3)^2 +\g^2 E_iE_i/2$. So one first
may solve the {\em unmodified} $su(2)$ Gauss law (on the classical as well as
on the quantum level) and then  is left with an ordinary
$R^2$-gravity theory as studied already before with an effective
cosmological constant $\L= - \g^2 E_iE_i/2\b^2$.  The coupling between the
gravity and the Yang-Mills system is thus seen to be  quite 'minimal', but of
course not zero.

From the point of view of the field content and the structure of the  action,
\re{S}) is an  obvious two-dimensional analogue of the gravity-Yang-Mills
system in four dimensions. From the technical point of view it is incomparably
simpler. This is precisely what one expects from a model to develop and/or test
conceptual ideas. Although the 'problem of time'  arises already in a  theory
of pure gravity as well, we have chosen to incorporate also the Yang-Mills part
in the action.  One of the reasons for doing so is that at any
point of the considerations we can  'turn off' the gravity curvature
by means of the limit $\b \to 0$. We are then basically left with a pure
Yang-Mills system. In its ordinary formulation the latter, however, has a
nonvanishing Hamiltonian and thus a meaningful Schr\"odinger equation, which
should be somehow reproduced in the gravity flat limit.  The coefficients in
\re{S}), where $\b$ and $\g$ are understood to be real, have been chosen so as
to avoid technical complications as far as possible: In particular there
will be no discrete indices within the wave functions arising from the gravity
sector, since $\L$ is effectively negative (cf.\ Fig.\ 1 and the discussion in
the previous sections).

In explicit terms the constraints following (naturally) from $L_H$ are
\ba
G_a&=& \6 B_a + \e^b{}_a B_b \o_1 + \e_{ab}[-\b^2(B_3)^2+\g^2tr E^2]   e_1{}^b
, \la{Ga}\\ G_3&=& \6 B_3 + \e_b{}^aB_a e_1{}^b,
\la{G3}
\ea
beside the unmodified $SU(2)$ Gauss law $G \approx 0$.  We will not attempt to
reformulate these constraints so as to possibly cure the global deficiencies of
them with respect to diffeomorphisms  noted at the end of the previous section.
Instead we proceed with a straightforward quantization.

There are two independent Dirac observables as functions of the momenta:
\ba q_1&=& {-1 \0 \pi} \oint tr(E^2) dx^1 \equiv {1 \0 2\pi}
\oint E_iE_i dx^1 \nn q_2 &=& {1 \0 2\pi}
\oint [(B)^2 -{2 \0 3} \b^2 (B_3)^3 +2\g^2 tr(E^2)B_3]dx^1 \, ,  \nonumber \ea
where $q_s \equiv \oint C^s dx^1/2\pi$ and $C^1$, $C^2$ are (the) two Casimir
functions of the target space Poisson structure.  The corresponding level
surfaces have topology $S^2 \times {\bf R}^2$ for $q_1 \neq 0$ and ${\bf R}^2$
for $q_1=0$.\footnote{Within the latter level surface the origin is an integral
surface by itself.  We will in the following disregard this small
complication.} This gives rise to the quantization condition (cf.\ end of sec.\
\ref{secquan}): $q_1 = n^2/4, \, n \in {\bf N}_0$.  Thus the physical wave
functions take the form
\be \Psi = \exp \left({i \0 \hbar} \oint (E_3 \6 \varphi \pm \ln B_\mp \, \6
B_3 dx^1\right) \, \Psi_0(n,q_2),  \,\, q_2 \in {\bf R},
\la{psi1} \ee
having written the phase factor in some local target space coordinates with
$\tan \varphi \equiv (E_2/E_1)$.

Expanding the physical  wave functionals in terms of eigenfunctions $\vert n
\>$ of $q_1$, we may write alternatively
\be \Psi= \sum_{n=0}^\infty
\exp \left({i \0 \hbar} \pm \oint  \ln B_\mp \, \6B_3 dx^1\right)
\ti \Psi_n(q_2) \vert n \> \,. \,\, \la{psi2} \ee
This makes contact with our previous observation: The coefficients in the above
expansion are $R^2$-gravity wave functions for the respective cosmological
constant $ - \g^2 n^2/8\b^2$.

The inner product with respect to $q_2$ is determined by the hermiticity
requirement on (cf.\ \re{P2}))
\be p_2 = - {1 \0 2} \oint {e_1{}^\pm \0 B_\mp}dx^1,   \la{p} \ee
the Dirac observable conjugate to $q_2$:  as noted already previously, $p_2$
acts as the usual derivative operator on $\ti \Psi_n$, thus leading to the
ordinary Lebesgue measure $dq_2$. Of course the hermiticity of $q_1$ leads to
the orthogonality of $\vert n \>$ for different $n$.

We end up with the Hilbert space ${\cal H}$ of an effective two-point particle
system with  nontrivial phase space topology (giving rise to the discrete
spectrum of $q_1$).  As a basic set of operators
acting in $\cal H$ we could use $q_2,p_2$, $q_1$, and $tr [{\cal P} \,\exp
(\oint A_1dx^1)]$. From the latter one may construct a ladder operator: $n
\to n + 1$.

All operators acting in $\cal H$ are thus found to be expressible in terms of
$q_2,p_2$, and the number and ladder operators. However, we do not have an
operator such as $g_{\m\n}(x^\m)$.  Following, furthermore, any  textbook on
elementary quantum mechanics, the next step in the quantization procedure would
be to introduce an evolution parameter 'time', which we will call $\t$, and to
require the wave functions to evolve in this parameter according to the
Schr\"odinger equation.  In the present case, however, the Hamiltonian following
from \re{SH}) is a combination of the constraints, \be H=-\oint [e_0{}^aG_a +
\o_0 G_3 +tr(A_0 \, G)], \la{H} \ee so that the naive Schr\"odinger equation
becomes meaningless.

Both of these items, the nonexistence of space-time dependent quantum operators
as well as the apparent lack of dynamics, are correlated and they are not just
a feature of the topological theory \re{S}). Also in four dimensional gravity
the quantum observables are some (not explicitly space-time dependent) holonomy
equivalence classes and the Hamiltonian vanishes when acting on physical wave
functions  \cite{Ash}. Diffeomorphisms are part of the symmetries of any
gravity theory; as a consequence the Lie derivative into any 'spatial'
direction can be found to equal  the  Hamiltonian vector field of some linear
combination of the constraints (in our case ${\cal L}_1 =e_1{}^aG_a +\o_1 G_3
+tr A_1 \, G$, cf.\ Eq.\ \re{diff})), whereas, on shell, $x^0$-diffeomorphisms
will be generated by the Hamiltonian $H$.  Thus, although 4D gravity has local
degrees of freedom, any of its (uncountably many)  Dirac observables will be
also space-time independent.

To orientate ourselves as of how to introduce  quantum dynamics within such a
system, let us have recourse to the simple case of a nonrelativistic particle
(NRP). As is well known, any Hamiltonian system can be reformulated in time
reparametrization invariant terms.  In the case of the NRP,
\be \int (p {dq \0 dt} - {p^2 \0 2})dt =\int (p \dot q  - {p^2 \0 2} \dot t)
d\t, \la{rep} \ee the equivalent system has canonical coordinates $(q,t;p,p_t)$
and the 'extended' Hamiltonian  is proportional (via a Lagrange multiplier) to
the constraint $K=p^2/2 + p_t \approx 0$.  Quantizing this system, e.g., in the
coordinate representation, we observe that the implementation  of the
constraint $K\psi(q,t)=0$ is equivalent to the Schr\"odinger equation of the
original formulation, if one reinterpretes the canonical variable $t$ as
evolution parameter $\t$. Therefore, given this formulation of the NRP or
similarly of any other system, the postulate of a Schr\"odinger equation within
the transition from the classical to the quantum system becomes superfluous;
rather it is already included within the Dirac quantization procedure in terms
of a constraint equation.

The identification $t =\t$ above can be looked upon also as a gauge condition
with gauge parameter $\t$.  This interpretation is helpful for the quantization
of the parametrization invariant NRP in the momentum representation
$\psi(p,p_t)$, in which case the space of physical wave functions is isomorphic
to the space of functions of the Dirac observable $p$. The gauge condition
$\bar K \equiv t-\t =0$ provides a perfect cross section for the flow of $K$.
Thus it is possible to determine any phase space variable in terms of the Dirac
observables $p$, $Q=q-pt$, as well as the gauge fixing parameter $\t$.
Interpreting $\t$ as a dynamical flow parameter 'time', the obtained evolution
equations for $p$ and $q$, transferred to the quantum level as $q(\t)=i \hbar
\, d/dp +   \t p$, $p(\t)=p$,  become equivalent to the Heisenberg evolution
equations of the parametrized  NRP.

The operator $q(\t)$ above  corresponds to a measuring device that determines
the place of the particle at time $\t$.  A measuring device that determines the
time $t$ at which the particle is at a given point $q=q_0$, on the other hand,
corresponds to the alternative gauge condition $\ti K \equiv q-q_0 =0$. $\ti K$
provides a good cross section only for $p\neq 0$. Ignoring this subtlety, e.g.
by regarding only wave functions with support at $p \neq 0$, the (hermitian)
quantum operator for such an experiment is $t(q_0)=-i \hbar \,[(1/p) d/dp
- (1/2p^2)] + q_0/p$.  In this second experimental setting Heisenberg's 'fourth
uncertainty relation' between time $t$ and energy $p^2/2 \sim -p_t$, usually
motivated only heuristically, becomes a strict mathematical equation. We learn
that different experimental settings are realized by means of different gauge
conditions, and, at least in principle, vice versa.

The wave functions of \re{SH}) are basically functions of the Dirac
observables, although  part of the latter  became discretized in the quantum
theory.  Transferring the ideas above to the gravity system, we should find
gauge conditions to the constraints \re{Ga}, \ref{G3}).  (It will not be
necessary to gauge fix also the $su(2)$ Gauss law $G$).  As such we will choose
\be \6 B_+=0, \quad B_3  + \t B_+ =0, \quad e_1{}^- =1.  \la{eich} \ee
It is somewhat cumbersome to convince oneself that this is
indeed a good gauge condition. However, for $q_1 \neq 0$  it
provides even a globally well-defined cross section.

[One possibility to check the obtainability of \re{eich}) is to carefully
analyze the corresponding Faddeev matrix, taking into account that the
constraints are not completely independent due to \re{depen}).  This (infinite
dimensional) matrix turns out to be nondegenerate, iff $B_+ \oint e_1{}^- dx^1
\neq 0$. For $q_1 \neq 0$ any gauge orbit in the loop space contains a
representative fulfilling this condition, which suffices to prove the assertion
since the space of gauge orbits is connected in the case under study (no
quantum number $n$).]

The gauge conditions \rp eich together with the constraints allow to express
all gravity phase space variables in terms of Dirac observables.  In this way
one obtains evolution equations such as
\be B_-(\t)=-{1 \0 2\pi} p_2q_2 -{\g^2 \0 2}  q_1 \t - {\b^2\pi^2
\0 3 (p_2)^2}\t^3 ,\quad B_+(\t)={-\pi \0 p_2}. \la{B-} \ee
Antisymmetrizing this with respect to $q_2$ and $p_2$,
\re{B-}) can be taken as an operator in the Hilbert space $\CH$
defined above.\footnote{The elementary procedure above coincides with the use
of Dirac brackets for $\t$-dependent systems (in which case one extends the
symplectic form by $d\t \wedge dp_\t$); this explains also $B_-$ and $B_+$ do
not commute anymore.} Similarly one finds  $g_{11}(x^0)=2e_1{}^+(x^0)= -p_2
B_-(x^0)/\pi$, $(x^0\equiv \t)$, which now, up to operator ambiguities, becomes
a well defined operator in our small quantum gravity theory, too.

Requiring that the $\t$-dependence of \re{eich}) is generated by the
Hamiltonian $H$, the gauge conditions determine also the zero components of the
zweibein and the spin connection.  Actually, one zero mode of these Lagrange
multiplier fields $e_0{}^a, \o_0$ remains arbitrary as a result of the linear
dependence \rp depen of the constraints $G_i$ (cf.\ also
\cite{p2}). Requiring this zero mode to vanish as a further gauge condition,
one finds $e_0{}^+=1$ and $e_0{}^-=\o_0=0$.  In other gauges the Lagrange
multipliers can become also nontrivial quantum operators. Furthermore, it is a
special feature of the chosen gauge (adapted to the Killing direction $B_3 =
const$) that the obtained operators are $x^1$-independent.  Again different
choices of gauge conditions are interpreted as corresponding to different types
of questions or measuring devices.

The alternative  procedure to reintroduce time within the quantum theory of the
parametrization invariant NRP was the direct implementation of the gauge within
the wave functions. For this it was decisive that the initially chosen
polarization of the wave functions, $\psi(q,t)$, contained the phase space
variable subject to the gauge.  To implement \re{eich}) analogously within the
gravity theory under consideration, we Fourier transform \re{psi2}), multiplied
by $\d[\6 C^2]$, with respect to $B_-(x^1)$. The result is
\begin{eqnarray}
\lefteqn{ \exp \left({i \0 \hbar} \oint [E_3 \6 \varphi +
{\6 B_+ B_3+ [{\b^2 \0 3} (B_3)^3 -\g^2 tr(E^2)B_3]e_1{}^- \0
B_+}] dx^1 \right)}
\hspace{6cm} &&
\nn
&& \Pi_{x^1}\left({const \0  B_+}\right) \hat
\Psi_n(p_2), \label{psiha} \end{eqnarray}
in which $\hat \Psi_n$ is the Fourier  transform of the ordinary function $\ti
\Psi_n$.

Eq. \re{psiha})   is in agreement with the general solution of the quantum
constraints in a $(B_+,B_3,e_1{}^-,E)$ representation, if we stick to the
operator ordering resulting from the Fourier transformation  of \re{qcons}).
Putting, on  the other hand, all derivative operators in the quantum
constraints to the right to start with, we again find no  quantum anomalies in
the constraint algebra. However, the latter operator ordering violates the
conditions \rp depen and thus leads inevitably to an empty kernel of the
constraints.\footnote{It would be interesting to see, if a similar mechanism is
responsible for the apparent lack of physical states in four-dimensional $N=1$
supergravity \cite{Page}.}

 In the gauge \re{eich})  the quantum wave functions \rp psi2 take the form
\be \Psi= \sum_n \exp{\left[{-i \0 \hbar} \left({\g^2n^2 \0 8}\t +{\b^2\pi^2 \0
3 p_2^2}\t^3 \right)\right]}  c_n(p_2) |n\rangle,
\la{time}  \ee  where we have absorbed the divergent factor of
\re{psiha}), being a function of $p_2$, into $c_n(p_2)$.

At this point it is worthwhile to perform the limit $\b \to 0$.  In some sense
\rp S with  $\b =0$  is the parametrization (i.e.\ diffeomorphism) invariant
formulation of the usual   Yang Mills theory on the cylinder (with rigid
Minkowski background metric). If we ignore the $p_2$ dependence of $c_n$ for a
moment, \re{time}) with $\b=0$ indeed coincides with the time evolution
generated by the (nonvanishing) Yang Mills Hamiltonian $-\g^2 \oint tr E^2 dx^1
\equiv  \g^2 \pi q_1$.  This agreement gives support to the method used to
derive \re{time}).

The reason for the $p_2$-dependence of $c_n$ is due to the fact that in the
formulation \re{SH}) with $\b=0$ the metric induced circumference of the
cylinder became a dynamical variable (on shell one has $p_2 \propto
\oint_{B_3=const} \sqrt{g_{11}} dx^1$). Within \re{eich}) one finds $-\oint G_+
\sim H$ to effectively implement the Schr\"odinger equation corresponding to
\re{time}).  The effective Hamiltonian acting on $c_n  |n\rangle$ is $-(\g^2/2)
\oint tr E^2 dx^1 - \b^2\pi^2 \t^2/p_2^2$.  Thus generically the above
procedure yields time dependent Hamiltonians.

In the case of the unparametrized  NRP the 'Heisenberg picture' and the
'Schr\"odinger picture' approach to introduce dynamics are obviously equivalent.
Straightforward equivalence of these two approaches was established also for
the gauge $X_+ =1, \, X_3=\t, \, \6 e_1{}^- =0$ in \cite{p2}.\footnote{There
the analysis was performed for the KV-model \re{KV}), but is valid in an
obvious way also here.} It is, however, not quite clear if or in how far the
same  is true also for the present incorporation  of the gauge conditions
\re{eich}). Further investigations into this direction, analyzing the subject
also from a more abstract point of view, are desirable.

The strategies developed at the example of a NRP to resolve the 'issue of time'
within a quantum theory of gravity produced, however, quite sensible results
for the toy model \re{S}). But they  relied heavily on either the knowledge of
all Dirac observables or on some specifically chosen polarization. To cope with
the considerable technical difficulties of a quantum theory of
four--dimensional gravity, it might be worthwhile to extend the applicability
of the method.

One way to do so within our model is to allow for equivalence classes of wave
functions coinciding at $\6 Q_2=0$, the latter condition being enforced  within
the inner product \cite{p2}.  In this way one can, e.g., implement the gauge
condition $\6 e_1{}^-=0$ as an operator condition in the $B$--polarization of
the wave functions as well. Still, however, a straightforward implementation of
$\oint e_1{}^-=const$  seems inadmissible also then in this polarization.

Given the open ends which may be found in this section, we still hope to have
convinced the reader that nontrivial (quantum) dynamics in a theory of gravity
corresponds, in one way or the other, to the choice of gauge conditions which
break the diffeomorphism invariance.

\chapter*{Figures}
\addcontentsline{toc}{chapter}{Figures}
\addtolength{\oddsidemargin}{-1cm}

\includegraphics{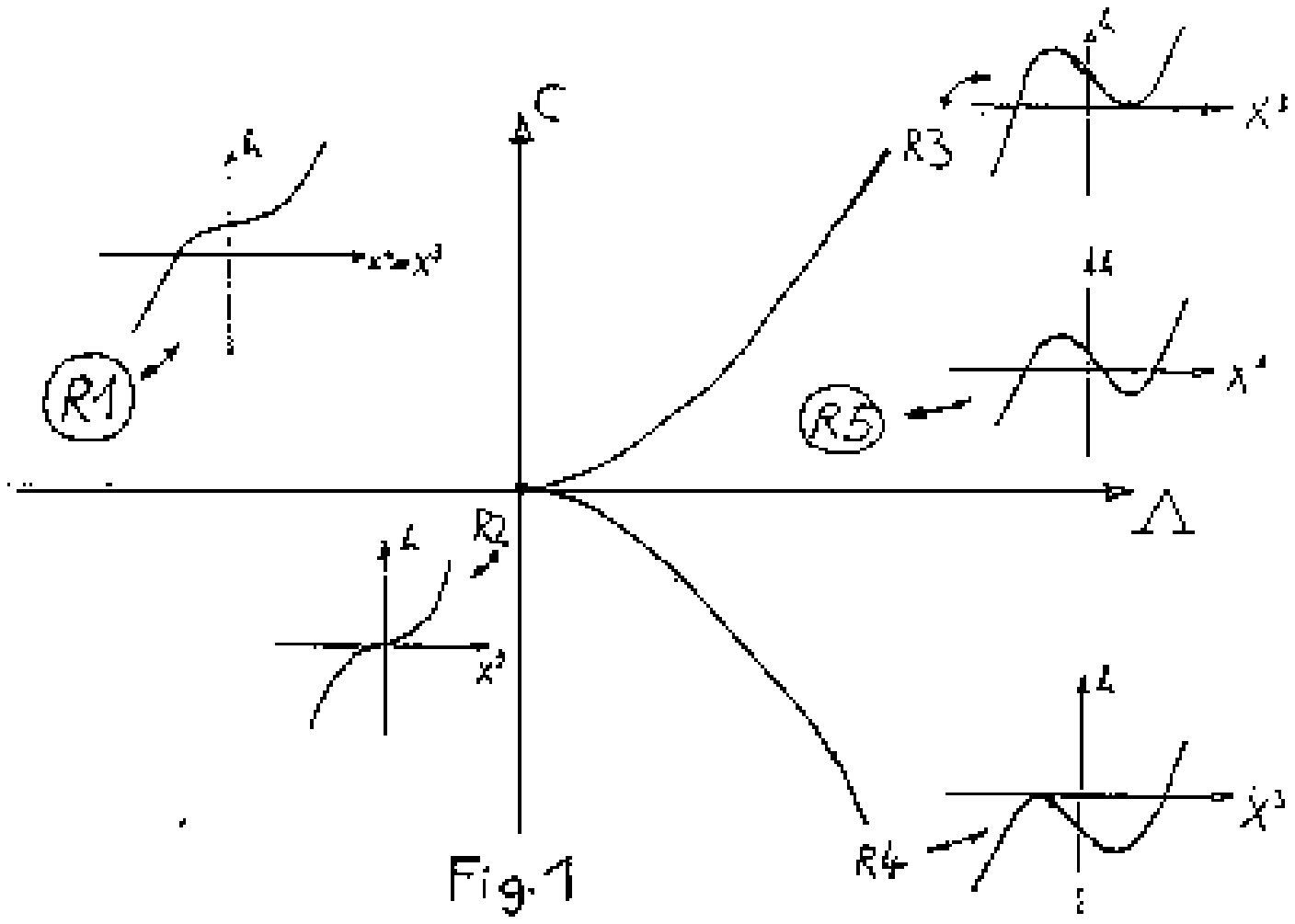}

\newpage
\addtolength{\oddsidemargin}{-2cm}
\includegraphics{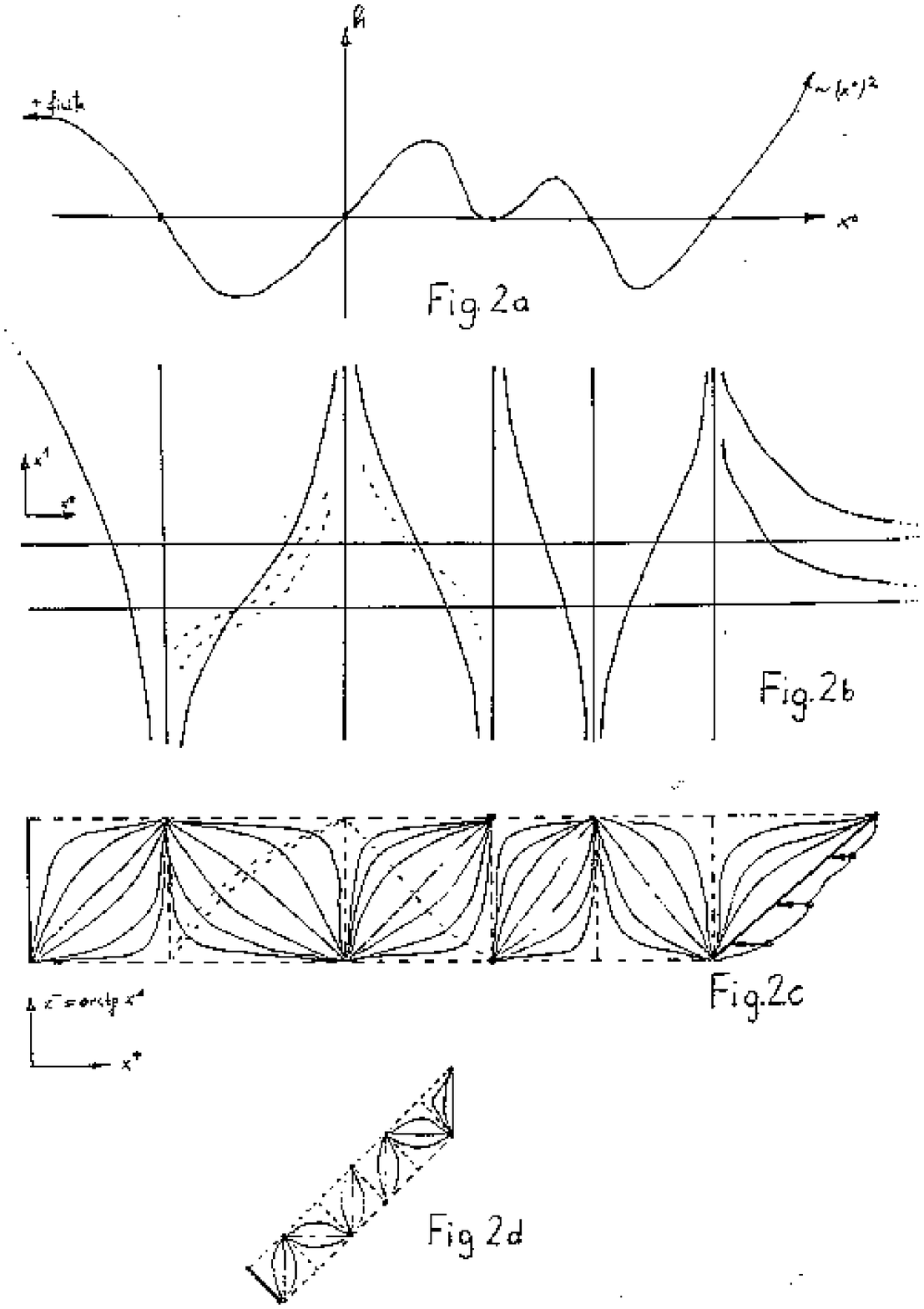}

\newpage
\addtolength{\textwidth}{4cm}
\includegraphics{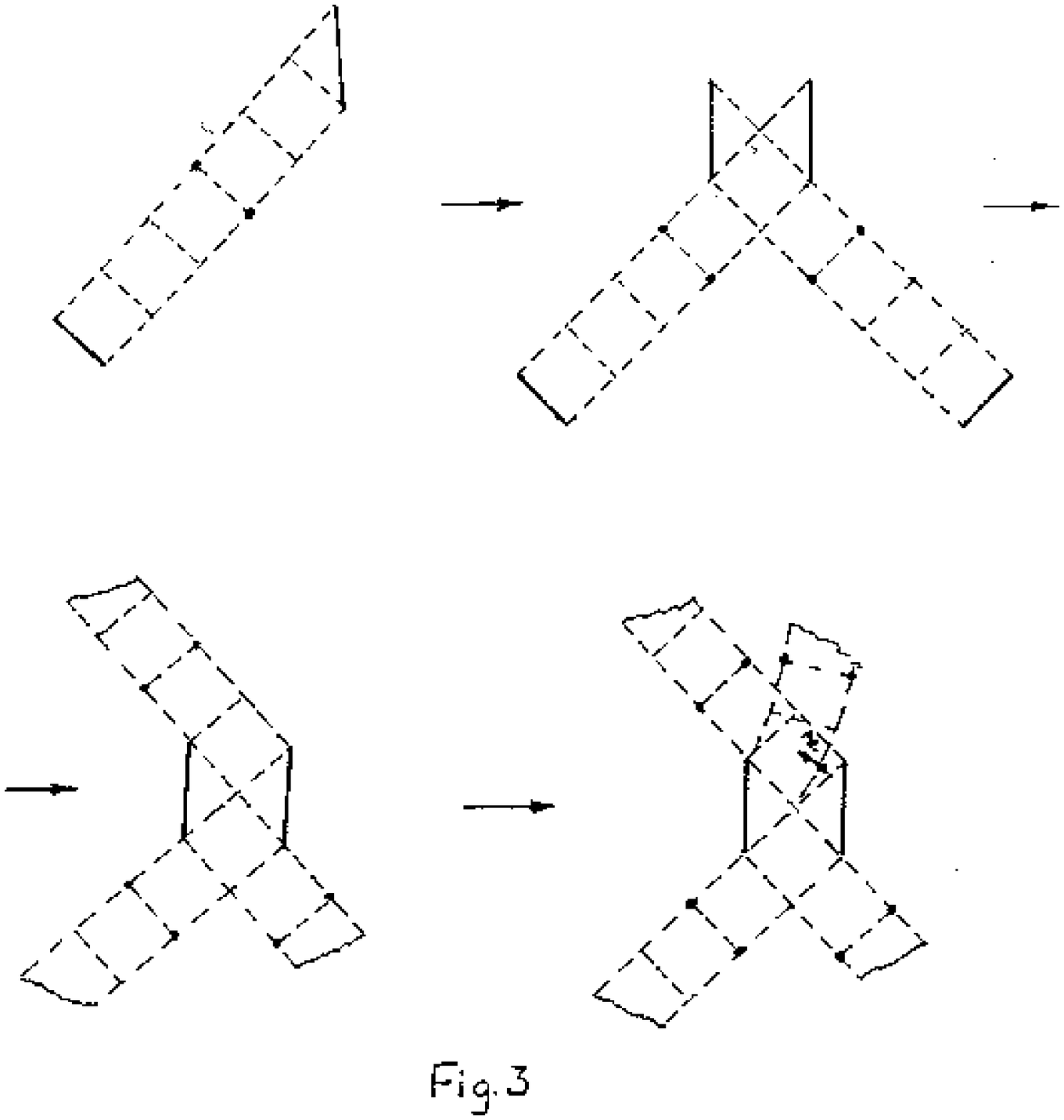}

\newpage
\addtolength{\oddsidemargin}{2.5cm}
\addtolength{\textwidth}{-5cm}
\includegraphics{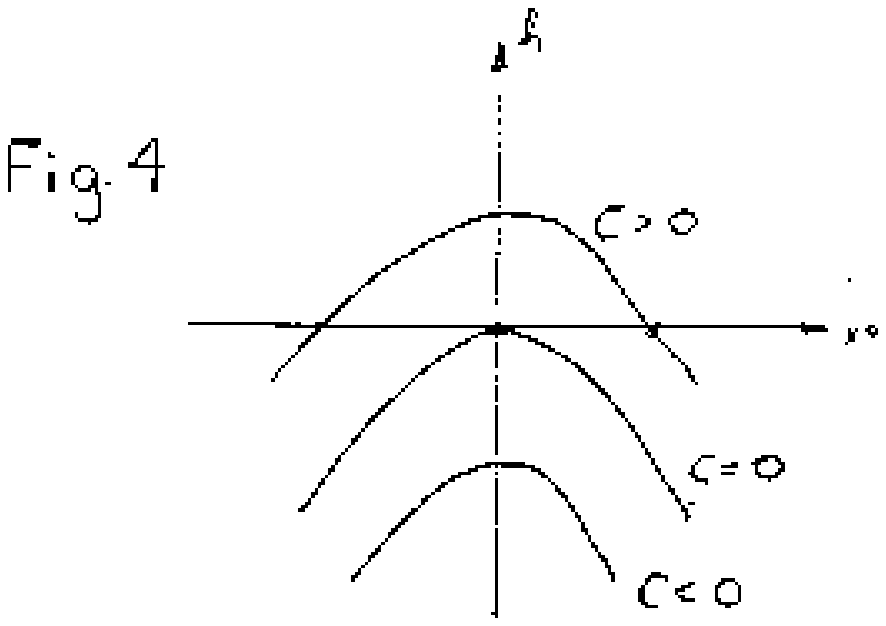}

\includegraphics{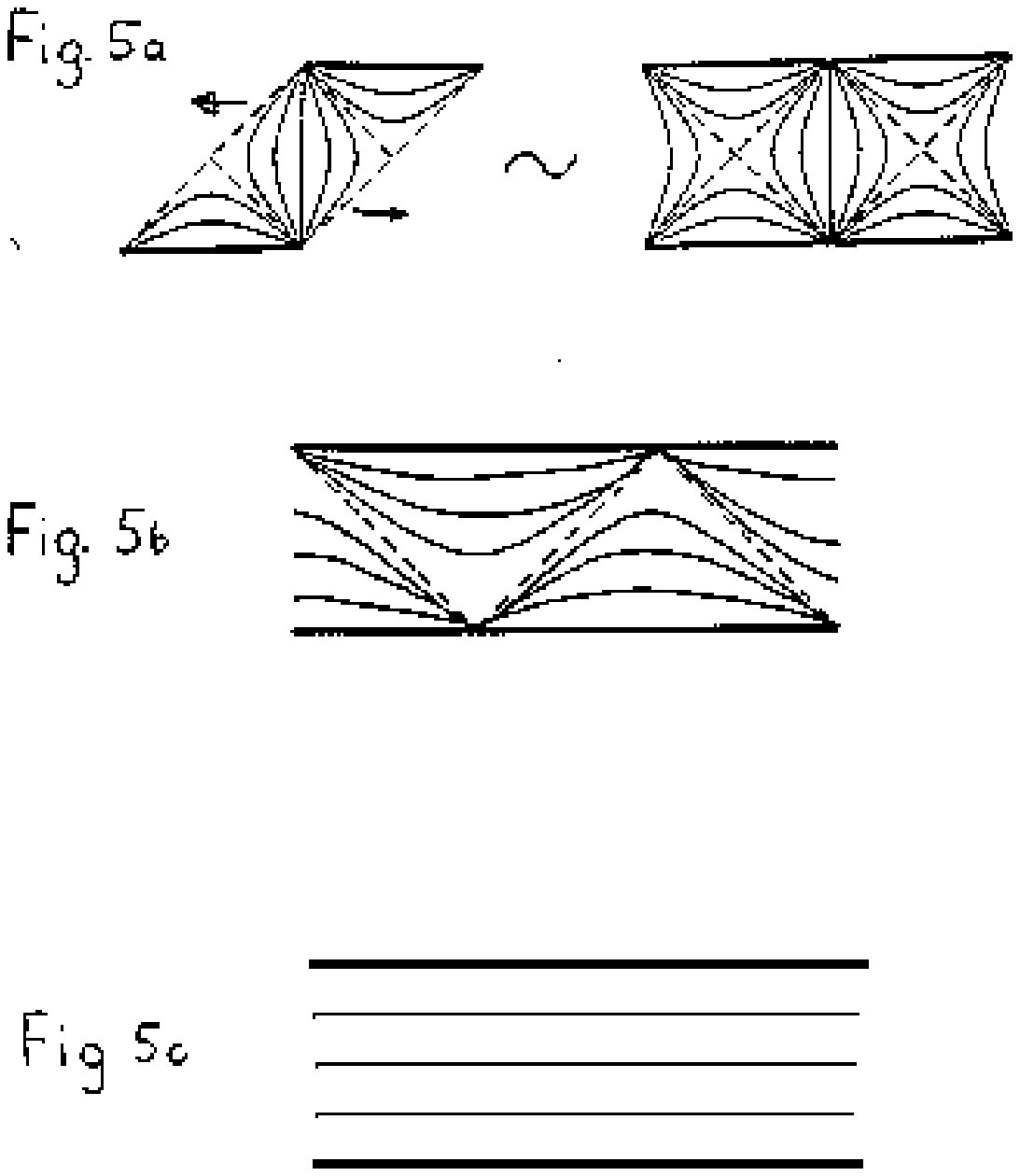}

\includegraphics{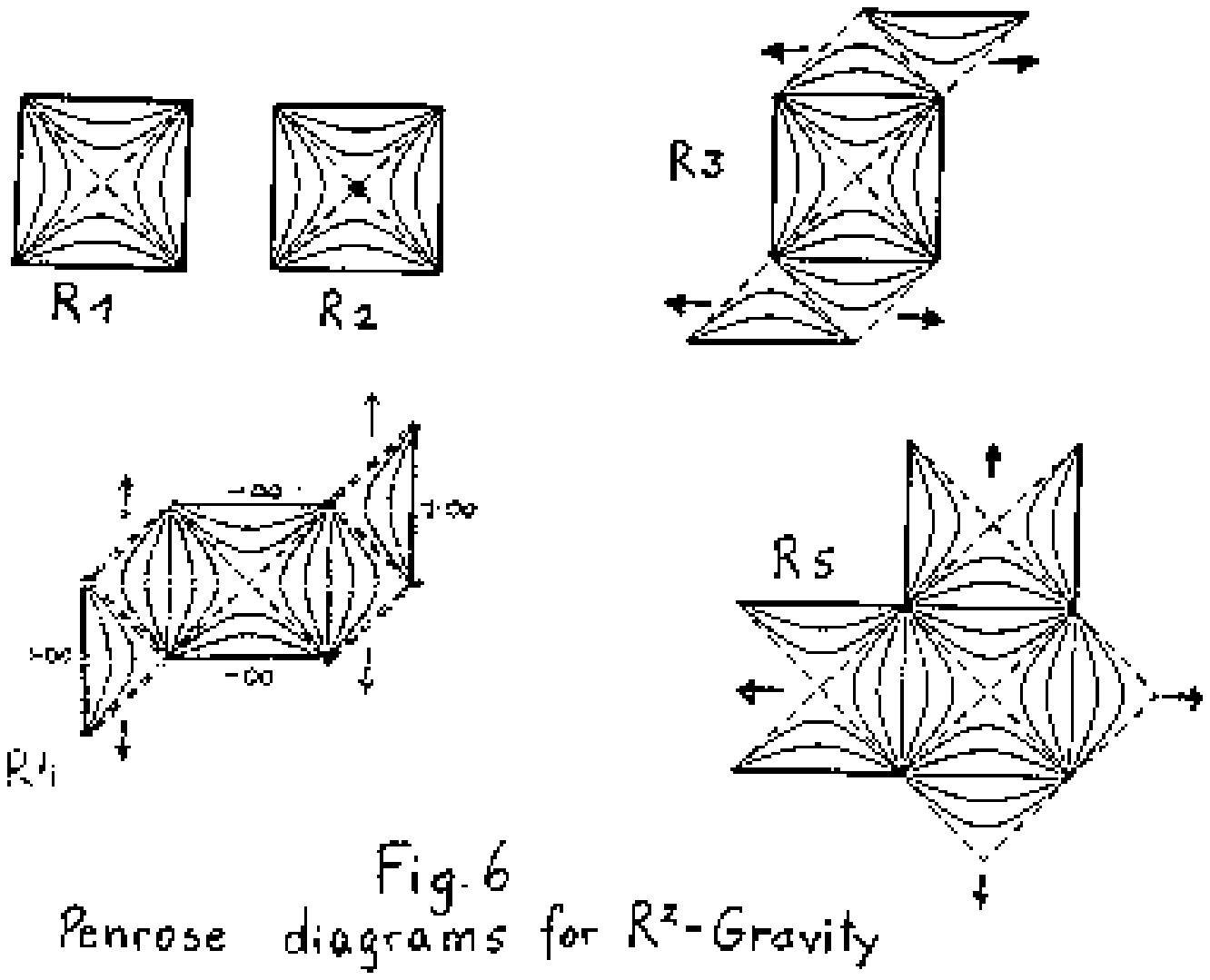}

\newpage
\addtolength{\oddsidemargin}{-2cm}
\addtolength{\textwidth}{4cm}
\includegraphics{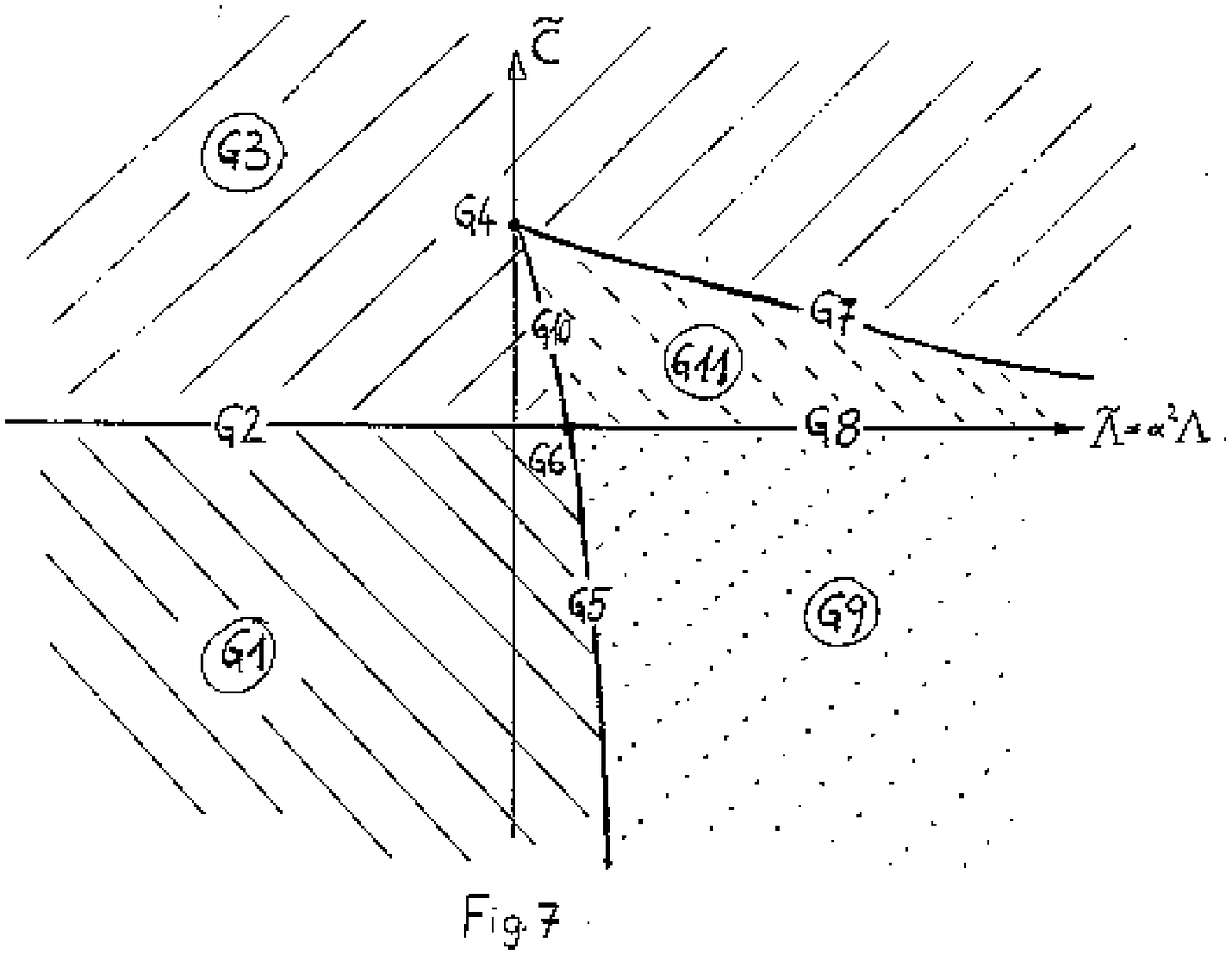}

\newpage
\addtolength{\oddsidemargin}{1cm}
\addtolength{\textwidth}{-2cm}
\includegraphics{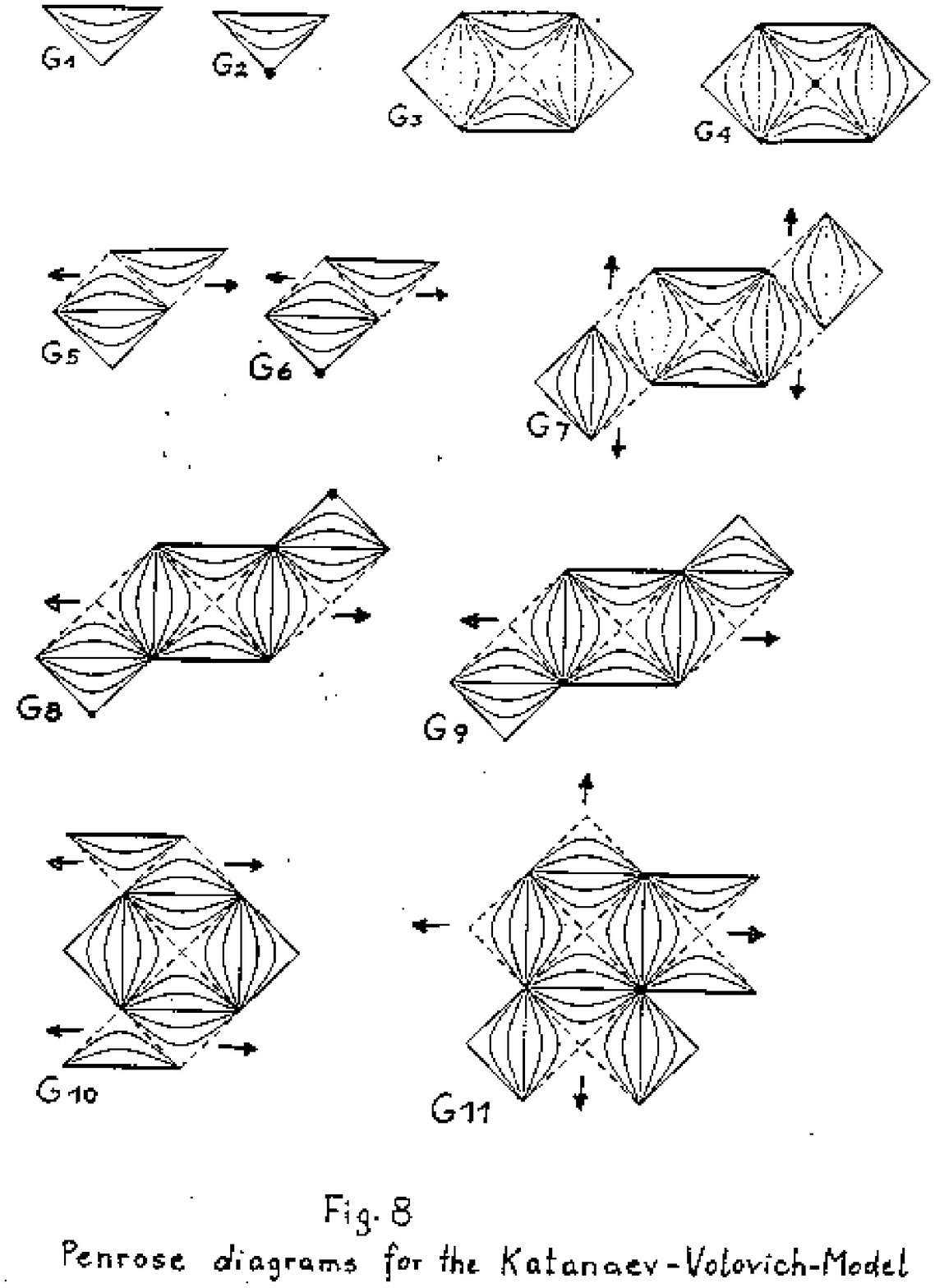}


\end{document}